\documentclass[onecolumn,amsmath,amssymb,nofootinbib,12pt]{article}
 
 \ifx\pdfoutput\not\undefined
 \pdfoutput=1
\fi
\usepackage{jheppub}
\usepackage{ifpdf}

\usepackage{mdframed}
\usepackage{comment}
\usepackage{graphicx,subcaption}
\graphicspath{ {figures/} }
\usepackage{epsfig}
\usepackage{pdfpages}
\usepackage{bbm}
\usepackage{graphicx,epstopdf}
\usepackage[numbers,sort&compress]{natbib}
\usepackage[makeroom]{cancel}
\usepackage{hyperref}
\hypersetup{pdftex,colorlinks=true,allcolors=blue}
\usepackage{array}
\usepackage[export]{adjustbox}
\usepackage{amsthm}
\usepackage[normalem]{ulem}

\theoremstyle{definition}

\usepackage[utf8]{inputenc}

\usepackage{tocloft}
\usepackage{bbold}
\usepackage{tikz}
\usetikzlibrary{backgrounds}
\usetikzlibrary{patterns}
\usetikzlibrary{arrows.meta}
\usetikzlibrary{shapes,snakes}
\tikzstyle{bag} = [align=center]
\usetikzlibrary{decorations.pathmorphing}
\def\bea{\begin{eqnarray}}
\def\eea{\end{eqnarray}}

\usepackage{hyperref}
\usepackage{subcaption}

 \newcommand{\badat}{\begin{alignedat}}
 \newcommand{\eadat}{\end{alignedat}}
 
 \def\be{\begin{equation}}
\def\ee{\end{equation}}

\def\p{\partial}

\usepackage{xcolor}

\newcommand{\pink}[1]{\textcolor{\pink}{#1}}

\definecolor{dblue}{rgb}{0.2,0.50,0.80}

\def\bz{{\bar z}}

\def\bz{{\bar z}}

\def\scri{\mathcal I}

\DeclareFontFamily{OT1}{pzc}{}
\DeclareFontShape{OT1}{pzc}{m}{it}{<-> s * [1.10] pzcmi7t}{}
\DeclareMathAlphabet{\mathpzc}{OT1}{pzc}{m}{it}

\definecolor{vert}{rgb}{0.1367 0.543 0.1367}

\title{Memory Correlators and Ward Identities in the `in-in' Formalism
 }

\author[a]{Ian Moult,}
\author[b]{Sruthi A. Narayanan,}
\author[b]{and Sabrina Pasterski}

\affiliation[a]{Department of Physics, Yale University, New Haven, CT 06511}
\affiliation[b]{Perimeter Institute for Theoretical Physics, Waterloo, ON N2L 2Y5, Canada}

\emailAdd{ian.moult@yale.edu}
\emailAdd{snarayanan@perimeterinstitute.ca}
\emailAdd{spasterski@perimeterinstitute.ca}

\date{\today}

\abstract{The symmetries of asymptotically flat spacetimes impose constraints on observables at infinity. The consequences of this have been extensively explored for $\mathcal{S}$-matrix elements, where soft theorems are known to be equivalent to Ward identities for asymptotic symmetries. However, recently there has been interest in broader classes of asymptotic observables. Here, we consider soft graviton insertions in the `in-in' formalism.  We derive a Ward identity for supertranslations and compute two point functions for the soft charges for `in-in' correlators. We find that the connected memory correlators are non-trivial in this set up and can be straightforwardly inferred from the average null energy (ANEC) correlators using observations from celestial Conformal Field Theory (cCFT).
}

\begin{document}

\maketitle

\section{Introduction}\label{sec:intro}

Asymptotic observables play a natural role in both collider experiments and quantum gravity in asymptotically flat spacetimes. Formally they are defined as operators inserted at the asymptotic boundary, which becomes a good approximation to collider experiments where the detector size is parametrically larger than the physical interaction scales. In quantum gravity, they play an even more important role, as such asymptotic observables are the only ones we expect to be well defined. Asymptotic symmetries, i.e. diffeomorphisms of asymptotically flat spacetimes that preserve the boundary structure, act non-trivially on these observables and should, in principle, constrain them. For scattering matrix ($\mathcal{S}$-matrix) elements this is understood as the observation that soft theorems are realizations of the Ward identities for these symmetries~\cite{Strominger:2013jfa,He:2014laa,Kapec:2014opa,Strominger:2017zoo}. 

However, $\mathcal{S}$-matrix elements are only one of a much larger class of asymptotic observables one can consider~\cite{Caron-Huot:2023vxl}. Indeed, in the study of asymptotic observables relevant to collider physics, the so called `in-in' formalism~\cite{Kosower:2018adc} is more commonly used. While traditional $\mathcal{S}$-matrix elements are calculated by inserting the relevant operator between an in-state ($|\rm{in}\rangle$) and an out-state ($|\rm{out}\rangle$), this formalism replaces $|\rm{out}\rangle$ by a copy of $|\rm{in}\rangle$. Since we are effectively tuning the incoming states and measuring the expectation values of calorimeter measurements, this can be thought of as an observable in the traditional sense.

There have been multiple recent efforts to understand the space of detector operators using insights from conformal field theory~\cite{Hofman:2008ar,Caron-Huot:2022eqs}. A central role is played by a subclass of detector operators with zero energy, such as the average null energy (ANEC) operator, constructed as integrals of certain components of the matter stress tensor $T_{\mu\nu}^M$, or similar operators constructed from integrals of conserved charges~\cite{Lee:2022uwt, Moult:2025nhu}. Precisely because these operators represent theoretical idealizations of calorimeter cells, they have a long history associated with the development of QCD \cite{Brandt1971,Christ1972,Sterman:1975xv,Basham1979}. After significant developments in their understanding in conformal field theories \cite{Hofman:2008ar,Belitsky:2013ofa,Belitsky:2013bja,Belitsky:2013xxa,Kologlu:2019mfz,Kologlu:2019bco}, they have been resurrected as practical observables for understanding the Standard Model at colliders \cite{Dixon:2019uzg,Chen:2020vvp,Lee:2022uwt,Chen:2022swd,Komiske:2022enw}. For a review, see \cite{Moult:2025nhu}. Curiously, these detector operators can be shown to obey the asymptotic symmetry algebras of asymptotically flat spacetimes in any unitary CFT with minimal assumptions~\cite{Cordova:2018ygx}.

When we couple to gravity, these operators become the so-called `hard charges' in the $\mathcal{S}$-matrix Ward identities mentioned above~\cite{Hu:2022txx}. These soft operators are also naturally boost eigenstates, making them amenable to the celestial conformal field theory (cCFT) description of a flat hologram~\cite{Raclariu:2021zjz,Pasterski:2021rjz,Pasterski:2021raf}. While a complete understanding of the  
operator dictionary for this flat hologram~\cite{Pasterski:2017kqt, Donnay:2018neh, 
Guevara:2024ixn,Kulp:2024scx} should be tantamount to cataloging the detector operators~\cite{
Caron-Huot:2022eqs,transverse_chang_2022,generalizing_korchemsky_2022}, and the bases seem amenable, there has been minimal cross talk between these two sub-fields. Some efforts to bridge this gap between the language of detector operators, cCFT and other soft physics results have been made in~\cite{Hu:2022txx,Hu:2023geb,Belin2019,lightray_gonzo_2021,Herrmann:2024yai}, but there is much more to be done.

The goal of this paper is to understand how the soft physics results are recast in the `in-in' formalism.\footnote{Curiously the cosmological analog of the soft-theorem/Ward identity story~\cite{Hinterbichler:2013dpa}
was developed around the same time as~\cite{Strominger:2013jfa} and employs an `in-in' formalism in their derivation. As will be discussed in more detail in~\cite{bjp} these adiabatic modes are not quite the same as the asymptotic symmetries, even in flat space settings~\cite{Mirbabayi:2016xvc}.} Not only is this another step towards making better contact between the celestial holography and detector operator/collider physics literature, it also provides us with some fruitful results. First, we will see that in deriving the `in-in' analog of the Ward identity, the soft charge insertions will be simply related to the corresponding ANEC insertions. This implies that the connected two point functions of these soft charges will be non-zero -- an observation which we would not readily observe from $\mathcal{S}$-matrix insertions~\cite{Himwich:2020rro,Pasterski:2021fjn,Pasterski:2021dqe} and which has potential implications for recent experimental proposals~\cite{He:2024vlp}. 
Furthermore, by combining these observations with results from cCFT that relate soft gravitons with different helicities, we will see that we can also evaluate memory correlators in terms of the ANEC correlators. This relationship would be rather non-trivial to derive if we had, instead, tried to compute the soft limit of the graviton waveform, as the phase space integrals naively take a different form from the ANEC correlators.

This paper is organized as follows. We start by reviewing some preliminaries in section~\ref{sec:prelim}. In particular, we review the original derivation of the supertranslation Ward identity for $\mathcal{S}$-matrix elements in~\ref{sec:softWard} as well as salient aspects of the `in-in' formalism and energy detector operators in~\ref{sec:inin}. 
We then turn to soft charge insertions in the `in-in' formalism in section~\ref{sec:softcharge}, deriving an `in-in' version of the supertranslation Ward identity in~\ref{sec:ininWard}, and computing the two point correlators in~\ref{sec:QSQS}. We then use the celestial diamond~\cite{Pasterski:2021fjn,Pasterski:2021dqe} framework to lift these to memory correlators in section~\ref{sec:memmem}, before closing with a discussion of our results and future directions in section~\ref{sec:disc}. 

While this work was in preparation~\cite{Gonzalez:2025ene} was posted, which contains some overlapping material.

\section{Preliminaries}\label{sec:prelim}
In this section we will set up some notation and review necessary background material for the rest of this work. This includes the derivation of the supertranslation Ward identity in the standard $\mathcal{S}$-matrix formulation in section~\ref{sec:softWard}, followed by a summary of perturbative calculations of `in-in' correlators in section~\ref{sec:inin}, and finally a discussion of how to recast the detector operators in~\cite{Herrmann:2024yai} into objects more familiar from the viewpoint of cCFT in section~\ref{sec:celenergy}.

\subsection{Soft Theorems as Ward Identities}\label{sec:softWard}
While it is known that each asymptotic symmetry has a corresponding Ward identity identifiable as a soft theorem, throughout this paper we will focus on the leading soft graviton mode, whose $\mathcal{S}$-matrix insertions have been shown to be equivalent to a Ward identity for supertranslations~\cite{Strominger:2013jfa,He:2014laa}. The goal of this section is to introduce enough of the steps in this derivation in order to understand how it fits into the `in-in' formalism outlined in section~\ref{sec:softcharge}. This will include: introducing a notion of soft and hard charges, how they relate to a term that can be antipodally matched, and how the soft theorem implies the Ward identity. We will follow the derivation in~\cite{He:2014laa} and restrict ourselves to massless scattering (see also~\cite{Strominger:2017zoo} for a nice pedagogical treatment). Most of our notation will follow \cite{He:2014laa}, with the small substitution of $a^{in}_\pm \mapsto a_\pm$ and $a^{out}_\pm \mapsto b_\pm$ to match the notation from~\cite{Caron-Huot:2023vxl} which we will use for the `in-in' formalism in the next section.\footnote{Note here that the $\pm$ subscripts denote the helicity of the particle, rather than whether it is incoming or outgoing. The $a,b$ differentiate in and out states.}

The essentials of the derivation are as follows. Bondi, van der Burg, Metzner, and Sachs identified the asymptotic symmetry group to include supertranslations~\cite{Bondi:1962px,Sachs:1962wk,Sachs:1962zza}. These symmetries independently translate different directions on the night sky. Namely, near future null infinity ($r\rightarrow\infty$, $u=t-r$ fixed) we have
\be
\xi^\mu \p_\mu \sim f(z,\bz)\p_u+...\,,
\ee
where $z=e^{i\phi}\tan\frac{\theta}{2}$ is a stereographic coordinate for the celestial sphere. From covariant phase space methods, one deduces that the charge generating these symmetries is related to the Bondi mass~\cite{Barnich:2009se,Barnich:2011ct,Barnich:2011mi}
\be
\mathcal{Q}^+=\frac{1}{4\pi G}\int_{\scri^+_-} d^2z \sqrt{\gamma} f m_B\,,
\ee
integrated at the past limit of future null infinity, denoted by $\mathcal{I}_-^+$. In what follows, $\gamma_{z\bz}$ will denote the metric on the unit round sphere in stereographic coordinates and $D_z$ is the corresponding covariant derivative. A key insight of Strominger was to ask if there exists a simultaneous transformation at past null infinity that would compensate this change, leading to the claim of a Ward identity of the following form
\be\label{eq:wardid}
\langle  \mbox{out}|\mathcal{Q}^+ \mathcal{S}-\mathcal{S}\mathcal{Q}^-|\mbox{in}\rangle =0.
\ee
This can be verified in the context of $\mathcal{S}$-matrix elements which we outline below.

The first thing to note is that this Ward identity is none other than an antipodal matching of the Bondi mass, $m_B$, across spatial infinity. However, to evaluate~\eqref{eq:wardid} we need to recast the Bondi mass in terms of the radiative data corresponding to the asymptotic states. The $uu$ component of Einstein's equations at large $r$ tells us that
\begin{equation}\label{eq:Guu}
\partial_u m_B=\frac{1}{4} D_z^2 N^{z z}+\frac{1}{4} D_{\bar{z}}^2 N^{\bar{z} \bar{z}}-4\pi G T_{u u}^{M(2)}-\frac{1}{4} N_{z z} N^{z z},
\end{equation}
where $N_{zz}=\p_u C_{zz}$ is the news, the shear, $C_{zz}$, is the radiative part of the metric and $T_{uu}^{M(2)}$ has already stripped out the $O(r^{-2})$ part of the matter stress tensor. We can recast the shear in terms of the asymptotic graviton creation and annihilation operators on the out state.\footnote{\label{ft:inoutconv} We're presenting the Ward identity as close as possible to the way it's normally written in the soft physics literature. After we clarify the in and out bases in the next section we will see that -- in contrast to~\cite{He:2014cra} -- it will behoove us to either tweak~\eqref{eq:wardid} or write~\eqref{eq:shear} in terms of incoming modes $a_\pm$, as we've done here. Namely, in this section, the out state would similarly be created with the $a$ modes, so that the Ward identity~\eqref{eq:wardid} has an explicit $\cal S$-matrix appearing.} The standard approach is to write the shear $C_{zz}$ as a mode expansion in terms of graviton creation and annihilation operators and then perform a saddle point approximation\footnote{Performing the saddle point approximation allows the exponential factors to be written solely in terms of $u$. It is also what defines $C_{zz}$ to be $\frac{1}{r}$ suppressed compared to the leading terms in the flat space metric.} to obtain
 \begin{equation}\label{eq:shear}
C_{z z}=-\frac{i \kappa}{4 \pi^2(1+z \bar{z})^2} \int_0^{\infty} d \omega_q\left[a_{+}\left(\omega_q \hat{x}\right) e^{-i \omega_q u}-a_{-}\left(\omega_q \hat{x}\right)^{\dagger} e^{i \omega_q u}\right]\,,
\end{equation}
where $\kappa^2=32\pi G$. By taking a $u$ integral of~\eqref{eq:Guu} and assuming, for simplicity, that $m_B$ vanishes at $\scri^\pm$ (which would be true for purely massless scattering, but can be generally relaxed) we see that the charges split into a sum of soft and hard terms 
\be\label{eq:ward2}
\langle  \mbox{out}|(\mathcal{Q}^+_S+\mathcal{Q}^+_H) \mathcal{S}-\mathcal{S}(\mathcal{Q}^-_S+\mathcal{Q}^-_H)|\mbox{in}\rangle =0\,,
\ee
where 
\be
\mathcal{Q}^+_S=-\frac{1}{16 \pi G} \int_{\mathcal{I}^{+}} d u d^2 z \sqrt{\gamma} f\left[D_z^2 N^{z z}+D_{\bar{z}}^2 N^{\bar{z} \bar{z}}\right],~~ \mathcal{Q}^+_H= \frac{1}{4\pi G}\int_{\mathcal{I}^{+}} d u d^2 z \sqrt{\gamma} f T_{u u}^{(2)}\,,
\ee
and similarly for the incoming states. We've grouped the last two terms in~\eqref{eq:Guu} to define $T_{uu}^{(2)}$ sans superscript.  We can recognize the hard charge as the shear inclusive ANEC operator\footnote{This can be done by writing out the light ray operators in the standard light-cone coordinates $x^\pm$ and then transforming those to Bondi coordinates.} while the soft charge involves derivatives of the memory operators 
\be\label{mem}
\mathcal{N}^{zz}:= \int du N^{zz}=\int du \p_u C^{zz}=\Delta C^{zz}\,,
\ee
and similarly for $N^{\bz\bz}$. Because the soft charge is linear in the metric, it will generate a constant shift. This is consistent with the interpretation that the supertranslations are asymptotic symmetries that are spontaneously broken by a choice of vacuum, unlike spacetime isometries. In particular, the Poincar\'e translations, a subset of the full set of supertranslation, obey $D_z^2 f=0$ and $\mathcal{Q}_S$ vanishes.

Explicitly verifying~\eqref{eq:ward2} amounts to evaluating both sides on $\mathcal{S}$-matrix elements. The hard charge just measures the energy carried by any external state while the $u$ integral of the news picks out a soft term in the shear. In particular one can show that 
\begin{equation}\label{eq:intNzz}
\mathcal{N}_{zz}=-\frac{\kappa}{4 \pi(1+z \bar{z})^2} \lim _{\omega \rightarrow 0^{+}}\left[\omega a_{+}(\omega \hat{x})+\omega a_{-}(\omega \hat{x})^{\dagger}\right]\,,
\end{equation}
or, equivalently, that the soft charge precisely extracts Weinberg's leading soft graviton theorem~\cite{Weinberg:1965nx} (see~\cite{Cachazo:2014fwa,Jorstad:2025qxu} for more on the (sub)subleading soft theorems). More explicitly, the way it works is that the soft theorem
\begin{equation}\label{eq:softfac}
\langle \mbox{out} | a_{ \pm}(q) S|\mbox{in} \rangle=S^{(0) \pm}\langle \mbox{out} |S|\mbox{in}\rangle+\ldots,~~~S^{(0) \pm}=\frac{\kappa}{2} \sum_i \eta_i \frac{\left(p_i \cdot \epsilon^{ \pm}\right)^2}{p_i \cdot q},
\end{equation}
where $\eta_i=\pm 1$ for outgoing/incoming particles, is effectively a Greens function for the differential operators in $\mathcal{Q}_S$, sourced by the energies of the external states. Namely
\be\label{eq:local}
D_z^2 \mathcal{N}^{zz}=D_\bz^2 \mathcal{N}^{\bar{z}\bar{z}}=4\pi G\sum_i \eta_i\omega_i{\delta^2(z-z_i)}/{\sqrt{\gamma}}
\ee
which we see by explicitly substituting in the expressions for the momenta and polarization tensors. 

Two comments are in order that will be useful in later sections of this work. First, the contributions from the different helicity sectors are the same. Second, an ingoing or outgoing soft graviton can couple to all legs while the ANEC operators acting on $|\mbox{in}\rangle$ or $|\mbox{out}\rangle$ only measure the energy deposited in $|\mbox{in}\rangle$ or $|\mbox{out}\rangle$, respectively. The former will rear its head again when we progress from soft charges to memory correlators in section~\ref{sec:memmem}, but will simplify some intermediate expressions along the way. The latter will be modified for the `in-in' version of the Ward identity in section~\ref{sec:ininWard}.

\subsection{The `in-in' Formalism}\label{sec:inin}

Scattering amplitudes traditionally describe the evolution between $|\rm{in}\rangle$ and $|\mbox{out}\rangle$. However, there are more asymptotic observables we can consider. Writing $\mathcal{S}=1+i\mathcal{T}$ where $\mathcal{S}^\dagger \mathcal{S}=\mathbb{1}$ in order to strip off the connected part, we know that inclusive cross sections measure projections of 
$\mathcal{T}^\dagger \mathcal{T}$
onto a fixed $|\rm{in}\rangle$ and then sum over $|\rm{out}\rangle$. On the other hand, the `in-in' formalism corresponds to inserting other asymptotic operators acting on $|\rm{out}\rangle$. It is essentially the Schwinger-Keldysh formalism~\cite{Keldysh:1964ud,Schwinger:1960qe} applied to correlation functions of operators placed at the asymptotic future, and thereby describes the measurements one would make in a particle detector. The method of computing these `in-in' correlators directly from on-shell scattering amplitudes is known as the Kosower-Maybee-O'Connell (KMOC) formalism~\cite{Kosower:2018adc}.

Since we want to consider objects of this form, we will review the setup for this class of asymptotic observables following the treatment in~\cite{Caron-Huot:2023vxl}.  Their assumptions are as follows:
\begin{enumerate}
\item The algebra of observables in the asymptotic past is generated by the creation and annihilation operators of stable particles 
with canonical commutation relations
\begin{equation}
[a_i,a_j^\dagger] = \delta_{i,j}2p_i^0(2\pi)^{3}\delta^{(3)}(\vec{p}_i-\vec{p}_j)
\end{equation}
where the subscripts label flavor and spin indices. 
\item An equivalent algebra exists for observables in the asymptotic future, denoted by $b_i,b^\dagger_i$ and these operators 
 are related by a unitary evolution operator $\mathcal{S}$:
\begin{equation}\label{eq:abbasis}
b_i = \mathcal{S}^\dagger a_i \mathcal{S}, \ \ b_i^\dagger =\mathcal{S}^\dagger a_i^\dagger \mathcal{S}.
\end{equation}
\item There exists a vacuum state for which
\begin{equation}\label{eq:abvac}
a_i|0\rangle = b_i|0\rangle = 0, \ \ \mathcal{S}|0\rangle = |0\rangle.
\end{equation}
\item While the stability of the external particles is expressed by the fact that the one particle states evolve trivially
\begin{equation}
b_i^\dagger|0\rangle = \mathcal{S}^\dagger a_i^\dagger|0\rangle = a_i^\dagger|0\rangle.
\end{equation}
\end{enumerate}
We can then build multi-particle in and out states as follows
\begin{equation}\label{eq:inout}
|m\cdots 1\rangle_{in} = a_m^\dagger\cdots a_1^\dagger|0\rangle, \ \ |m\cdots 1\rangle_{out} = b_m^\dagger\cdots b_1^\dagger|0\rangle
\end{equation}
so that an $\mathcal{S}$-matrix element can be expressed as 
\begin{eqnarray}
{}_{out}\langle m+n\cdots m+1|m\cdots 1\rangle_{in} & = & \langle 0|b_{m+n}\cdots b_{m+1}a^\dagger_{m}\cdots a^\dagger_1|0\rangle\cr
& = & \langle 0| a_{m+1}\cdots a_{m+1} \mathcal{S}a_{m}^\dagger\cdots a_1^\dagger|0\rangle\cr
& = & _{in}\langle m+n\cdots m+1|\mathcal{S}|m\cdots 1\rangle_{in}.
\end{eqnarray}
To get from the first to the second line we've used~\eqref{eq:abbasis} and~\eqref{eq:abvac} to conclude
\be
\langle 0|\prod_i b_i=\langle 0| \mathcal{S}^\dagger a_{m+n} \mathcal{S}\mathcal{S}^\dagger a_{m+n-1} \mathcal{S}...=\langle 0|\mathcal{S}^\dagger\prod_i a_i \mathcal{S}=\langle 0|\prod_i a_i \mathcal{S}.
\ee
Let us emphasize that this notation for $|\rm{out}\rangle$ is different from what commonly appears in the celestial literature (as in section~\ref{sec:softWard}) and we will switch over to this notation hereafter. In what follows, we will use these rules to explicitly write out the relevant asymptotic observables in terms of creation and annihilation operators acting on $|\rm{in}\rangle$ and $|\rm{out}\rangle$. 

In particular, the `in-in' observables we will be interested in will be of the form 
\be\badat{3}
\langle \mbox{in}'|(\prod_{i\in I} b_i^\dagger)(\prod_{j\in J} b_j) |\mbox{in}\rangle&= \sum_{out} \langle \mbox{in}'|(\prod_{i\in I} b_i^\dagger)|\mbox{out}\rangle\langle \mbox{out}|(\prod_{j\in J} b_j) |\mbox{in}\rangle\\
&=\sum_{out}\langle \mbox{in}'|\mbox{out},I\rangle\langle \mbox{out},J|\mbox{in}\rangle 
\eadat\ee
for some set of modes $I,J$. Here $|\mbox{in}\rangle$ is a generic in state of the form~\eqref{eq:inout}, and we will be interested in the limit where $|\mbox{in}'\rangle \mapsto |\mbox{in}\rangle$ so that this has an interpretation as an expectation value of future asymptotic operators in the state $|\mbox{in}\rangle$. We see that by inserting a complete set of $|\rm{out}\rangle$ states we can recast everything in terms of matrix elements of $\mathcal{S}$ or $\mathcal{S}^\dagger$. In principle the sum $|\rm{out}\rangle\langle \rm{out}|$ includes all possible multi-particle states, generating a perturbative expansion. In our explicit computations here as well as those done in~\cite{Henn:2019gkr,Herrmann:2024yai} we will choose to work with the lowest order observable of interest, thereby bypassing the need for a sum over all $|\rm{out}\rangle$.
\begin{figure}[htb]
    \centering
    \includegraphics[width=0.9\linewidth]{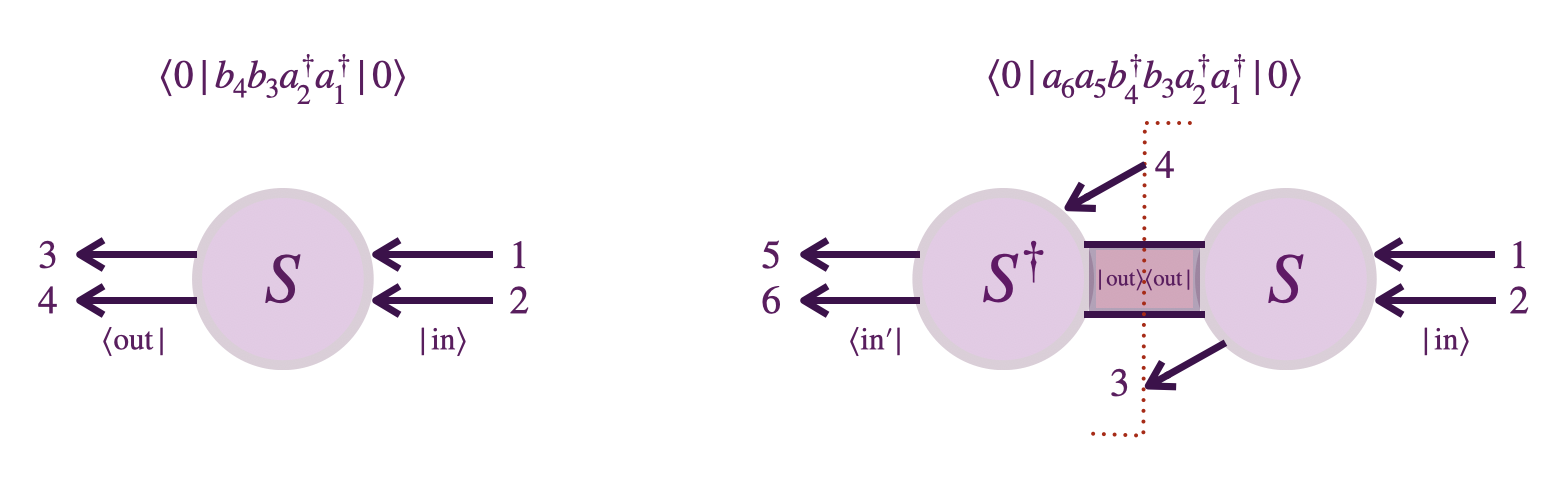}
    \caption{
    In contrast to $\mathcal{S}$-matrix elements (left), `in-in' correlators (right) involve inserting operators in the out state and summing over a complete set of intermediate states ($|\rm{out}\rangle\langle \rm{out}|$) on the positive-energy cut (dotted line).}
    \label{fig:inin}
\end{figure}

The diagrammatic relationship between `in-in' correlators and $\mathcal{S}$-matrix elements is illustrated in figure~\ref{fig:inin}. We can see that the `in-in' correlators are heuristically computed by gluing two standard scattering diagrams end to end. To compute these quantities we  apply the following rules. Following the notation in~\cite{Caron-Huot:2023vxl} we will use type I and type II to refer to components of the diagrams involving the $\mathcal{S}$ and $\mathcal{S}^\dagger$ contributions, respectively.  Then for all of our `in-in' observables we can compute each of the $\mathcal{S}$ and $\mathcal{S}^\dagger$ diagrams in perturbation theory. The usual Feynman rules\footnote{More often than not, the parts of the diagrams computed using normal Feynman rules are suppressed and lumped into part of the correlator that comes from the normal scattering amplitude.} are then supplemented by the following \label{pg:rules2}
\begin{enumerate}
    \item The external $|\mbox{in}\rangle$ and $\langle \mbox{in}'|$ legs will be truncated, any propagator between two type-I vertices will be a usual Feynman propagator.
    \item Any propagator between two type II vertices will be an anti-Feynman propagator. Each type II vertex and propagator should get an overall minus sign compared to its type-I counterpart.
    \item All of the states appearing in our $|\rm{out}\rangle$ sum will correspond to propagators between type-I and type-II vertices and the Feynman propagator should be replaced with
    \be
    \frac{1}{p^2+m^2-i\epsilon }\mapsto 2\pi i\theta(p^0)\delta(p^2+m^2).
    \ee
\end{enumerate}  
With these modified rules, the sum over $|\rm{out}\rangle$ is recast as if we are doing loop integrals with modified propagators along the positive-energy cut between the type-I and type-II vertices. Namely  
\be
\sum_{out} =\sum_{I} \prod_{i\in I} \int \frac{d^4 p_i}{(2\pi)^4}2\pi \theta(p_i^0)\delta(p^2_i+m^2_i).
\ee
For the `in-in' observables we are interested in here, we only need to consider situations with a single time-fold, whereas~\cite{Caron-Huot:2023vxl} allows for multiple time-folds.

\subsection{Celestial Energy Operators} \label{sec:celenergy}
We will now turn to how to evaluate a specific class of detector operators in the `in-in' formalism. We will start by reviewing the set up from~\cite{Caron-Huot:2022eqs,Herrmann:2024yai}. The perturbative computations in~\cite{Herrmann:2024yai} use the `in-in' formalism as reviewed in the previous subsection, and part of our goal here will be to present things in a way that will be be amenable for comparing to the Ward identity computations from section~\ref{sec:softWard}.

While more general detector operators have been considered in~\cite{Korchemsky:2021okt}, here we will focus on a class of detectors $\mathcal{D}_{J_L}(q)$ that measure the energy weighted to some power. Namely, for an asymptotic on-shell state labeled by $|X\rangle = |p_1,\cdots, p_n\rangle$ with $p_i = \omega_i(1,\hat{p}_i)$ being the momenta of each particle, the detector operator acts as\footnote{Note that we write the detector operator in terms of $q$ whereas in~\cite{Herrmann:2024yai} they use $z$. We have opted to change the notation here since we usually use $z,\bar{z}$ to denote celestial coordinates. Also note that, as compared to~\cite{Caron-Huot:2022eqs}, we have $J = 3-d-J_L$.}
\begin{equation} \label{eq:detector}
\mathcal{D}_{J_L}(q)|X\rangle = \sum_{i\in X}\omega_i^{2-d-J_L}\delta^{d-2}\left(\Omega_{\hat{q}}-\Omega_{\hat{p}_i}\right)|X\rangle\,,
\end{equation}
where $\hat{q}$ is a null direction associated to the vector $q = (1,\hat{q})$ and $\Omega_{\hat{q}}$ denotes the angular variables on a sphere. These are sometimes referred to as ``energy fluxes''~\cite{Caron-Huot:2022eqs} or ``energy-to-some-power'' operators~\cite{Herrmann:2024yai} to distinguish from the special case of the ANEC where $J_L=0$. Here we will adopt the term {\it celestial energy operators} to emphasize that these have definite weights under the boosts defining the celestial conformal dimension~\cite{Hu:2022txx} 
\be 
\Delta_{cCFT}=-J_L,~J_{cCFT}=0.  
\ee
An $n$-point function of detector operators is given in the `in-in' formalism as
\begin{equation}
\mathbb{E}_n \equiv \langle \mbox{in}|\mathcal{D}_{J_{L_1}}(q_1)\cdots\mathcal{D}_{J_{L_n}}(q_n)|\mbox{in}\rangle.
\end{equation}
As noted in the previous subsection, we sandwich a complete set of states in order to compute this quantity in terms of amplitudes, which are a more familiar observable. This complete set of states will generally include all possible multi-particle states. As outlined in detail in~\cite{Herrmann:2024yai}, the number of particles in $|\rm{out}\rangle$ and the number of detectors will determine the complexity of the computation. For instance, when there is one detector, considering a single particle in $|\rm{out}\rangle$ is relatively easy to compute but if there are two particles in $|\rm{out}\rangle$, there will be a loop integral to do over the extra momentum.  These contributions then contribute at different orders in the weak coupling expansion. Since we will not be summing over a complete set of states, we will denote the order of the computed correlator by a superscript $\mathbb{E}_n^{(i)}$. Going forward, we assume that $i=0$ denotes the lowest order contribution to the correlation function.\footnote{It is important to note that $i$ does not correspond to the number of particles in the out state.}

It is easier to first define the energy operators as acting on a state and then define the associated vertex function used in Feynman diagrammatic computations. Ideally, we would like to write all of these objects in terms of quantities more familiar in the celestial literature,  so let's massage~\eqref{eq:detector} a bit. Equation~\eqref{eq:detector} can be expressed in a covariant way as 
\begin{equation}
\mathcal{D}_{J_L}(q)|X\rangle = \sum_{i\in X}\int_0^\infty d\beta \beta^{1-J_L}\frac{\delta^{(d)}(p_i-\beta q)}{2\delta(q^2)}|X\rangle.
\end{equation} 
Noting that $\delta(q^2) = \frac{1}{2q^0}\delta(q^0-|q|)$ we see that 
\begin{equation}
\frac{\delta^{(d)}(p_i-\beta q)}{\delta(q^2)} = \frac{2}{\beta}\delta^{(d-1)}(\omega_i\hat{p}_i-\beta \hat{q}).
\end{equation}
Going forward, we specialize to $d=4$ and note that the the three-momenta can be written as 
\begin{eqnarray}
\hat{p} = \left(\frac{w+\bar{w}}{1+w\bar{w}}, -i\frac{(w-\bar{w})}{1+w\bar{w}}, \frac{1-w\bar{w}}{1+w\bar{w}}\right).
\end{eqnarray}
This allows us to expand the $d-1=3$ dimensional delta function as follows
\begin{equation}
\delta^{(3)}(\omega_i\hat{p}_i(w,\bar{w})-\beta\hat{q}(z,\bar{z}))  =  \frac{(1+w_i\bar{w}_i)^2}{2\omega_i^2}\delta\left(\omega_i-\beta\right)\delta\left(\vec{w}_i-\vec{z}\right)\equiv\frac{\delta\left(\omega_i-\beta\right)\delta\left(\vec{w}_i-\vec{z}\right)}{\sqrt{\gamma}\omega_i^2}
\end{equation}
where $\vec{z} = (z,\bar{z})$ and $\delta(\vec{z}_{ij}) = \delta(z_{ij})\delta(\bar{z}_{ij})$ is understood to be the two dimension delta function in what follows. This allows us to simplify the covariant expression to 
\begin{equation}
\mathcal{D}_{J_L}(q)|X\rangle = \frac{1}{\sqrt{\gamma}}\sum_{i\in X} \omega_i^{-J_L-2}\delta\left(\vec{w}_i-\vec{z}\right)|X\rangle.
\end{equation}
We see that when $J_L=-3$, this gives one power of the energy which is what we expect from the ANEC. In the context of amplitudes and the normal Feynman diagram computations, whenever there is a detector placed on a cut, we can insert the following vertex function
\begin{equation}
V_{J_L}(p_1,p_2) = \int_0^\infty d\beta \beta^{1-J_L}\frac{\delta^{(d)}(p_2-\beta p_1)}{2\delta(p_1^2)}.
\end{equation}
Here $p_1,p_2$ are the momenta coming in and going out of the detector. In the same way as above, we can write these in terms of celestial variables in $d=4$ and for $J_L=-3$
\begin{equation}
V_{J_L=-3}(p_1,p_2) = \frac{1}{\sqrt{\gamma}}\omega_2\delta(\vec{w}_{12}).
\end{equation}
In what follows, we denote $V_{J_L=-3}\equiv V$.

\begin{figure}[htb!]
    \centering
    \includegraphics[width=0.6\linewidth]{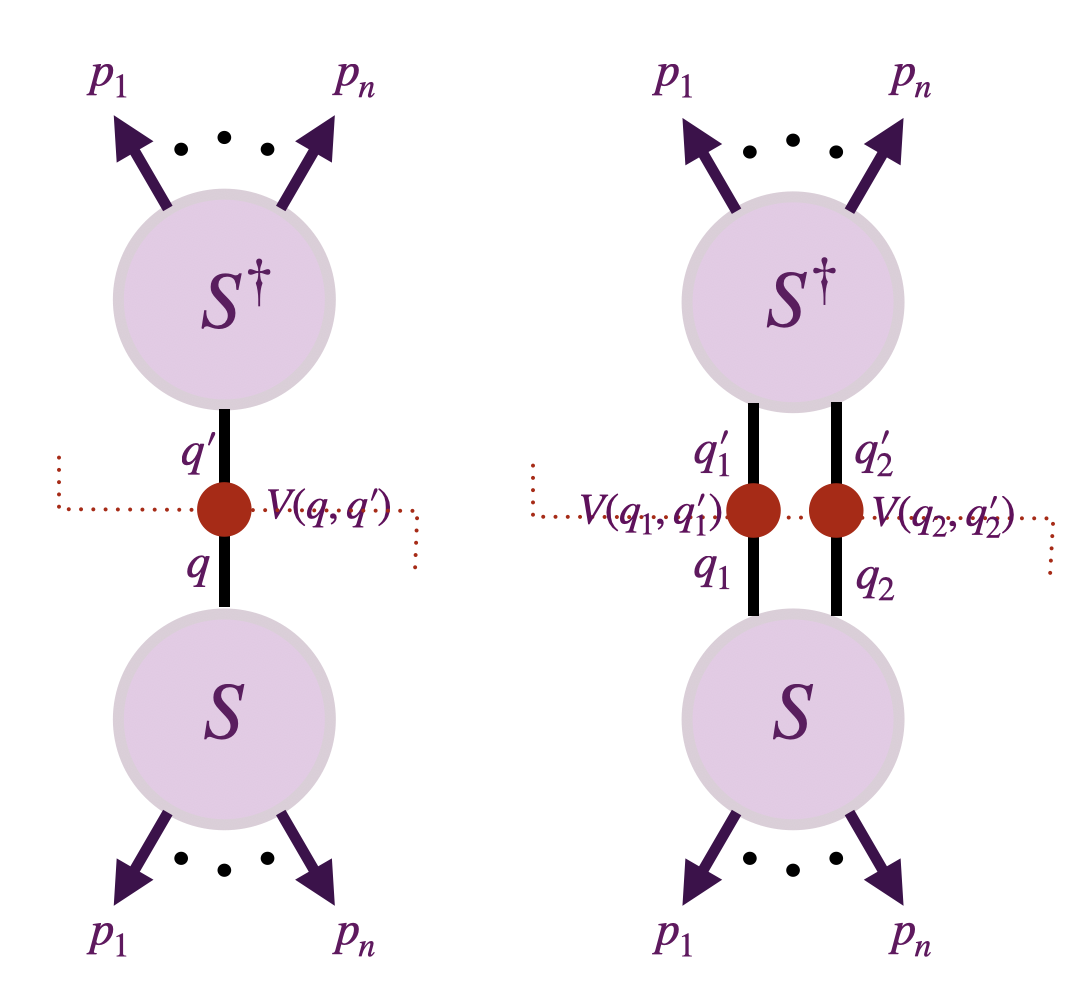}
    \caption{Diagrams in the `in-in' formalism that we use to compute the one-point and two-point functions. The red circles denote detectors that are at the cut and are represented by vertex functions $V(q,q')$ in the Feynman diagrams.} 
    \label{fig:one_point}
\end{figure}

\paragraph{One- and two-point functions of ANECs:} We are now in a position to compute the one- and two-point function of the ANEC operators given above. We will compute both of these at lowest order. Having these expressions in hand will make it easy to relate them to the soft charge correlators that we aim to compute. The one-point correlator of an energy operator $\mathcal{D}_{J_L}$ is given by 
\begin{equation}
\mathbb{E}_1 = \langle \mbox{in}|\mathcal{D}_{J_L}(q')|\mbox{in}\rangle.
\end{equation}
As we mentioned, we can compute this one-point function at different orders that each correspond to a different number of states at the cut. The first of these is represented in the first diagram in Figure~\ref{fig:one_point}. If we let the prepared $|\rm{in}\rangle$ contain $n$ particles of momenta $p_1\cdots p_n$ and also let $P = p_1+\cdots +p_n$, then the first order term that contributes to this one-point function is where there is a single particle in $|\rm{out}\rangle$. On one side of the cut, that state has momentum $q$, then it interacts with the detector and it's resulting momentum is $q'$. Writing this out in terms of the vertex and a single phase space integral gives 
\begin{equation}
\mathbb{E}_1^{(0)}(q'(\vec{w})) = \int d^4 q V(q,q')\left|\mathcal{M}_{n\rightarrow 1}^{(0)}(\{p_n\},q)\right|^2 \delta^{(4)}(q-P) .
\end{equation}
This is the leading order connected contribution to the ANEC one-point function where $|\mathcal{M}_{n\rightarrow 1}^{(0)}(\{p_n\},q)|^2$ denotes the squared $n\rightarrow 1$ amplitude at tree level. We see that the phase space integral enforces the momentum conservation. Of course, the existence of such an amplitude will depend on the theory. In the case where such an amplitude does not exist, the lowest order contribution to this one point function will be with more particles in the summed over out states. However, in order to keep this discussion clean and general, we will not specify a theory here and continue under the assumption of restricted kinematics as in~\cite{Herrmann:2024yai}.

We can take the one-point function contribution written above and plug in the vertex to write it in a form that will be more recognizable from the celestial point of view
\begin{eqnarray}\label{eqn:leadingone}
\mathbb{E}_1^{(0)}(\vec{w}) & = & \frac{1}{\sqrt{\gamma}}\int d^4 q\omega\delta(\vec{w}-\vec{w}')\left|\mathcal{M}_{n\rightarrow 1}^{(0)}(\{p_n\},q)\right|^2\delta^{(4)}(q-P).
\end{eqnarray}
For our purposes we will also need the two point function at lowest order which was not explicitly computed in~\cite{Herrmann:2024yai}. For this one, we use the second diagram in Figure~\ref{fig:one_point}. In terms of the vertex functions it is given by 
\begin{eqnarray}\label{eqn:leadingtwo}
& & \mathbb{E}_2^{(0)}(\vec{w}_1',\vec{w}_2')  = \int d^4q_1 d^4 q_2 V(q_1,q_1')V(q_2,q_2')\left|\mathcal{M}_{n\rightarrow 2}^{(0)}(\{p_n\},q_1,q_2)\right|^2\delta^{(4)}(q_1+q_2-P)\cr
& = & \int \frac{d^4q_1 d^4q_2}{\gamma} \omega_1\omega_2\delta(\vec{w}_1-\vec{w}_1')\delta(\vec{w}_2-\vec{w}_2')\left|\mathcal{M}_{n\rightarrow 2}^{(0)}(\{p_n\},q_1,q_2)\right|^2\delta^{(4)}(q_1+q_2-P).
\end{eqnarray}
We will use these expressions to relate to the one- and two-point functions of the soft charges.

\section{Soft Charge Insertions}\label{sec:softcharge}
We are now set up to examine how soft charge insertions in the asymptotic states are related to hard charge, namely ANEC, insertions.  As alluded to in the introduction, the latter have been studied extensively (see~\cite{Moult:2025nhu} for a pedagogical review). We will start by deriving an analog of the $\mathcal{S}$-matrix Ward identity from section~\ref{sec:softWard} in the `in-in' formalism.

\subsection{The `in-in' Ward Identity}\label{sec:ininWard}
We already saw that the leading soft graviton theorem encodes a Ward identity for the action of supertranslations on the $\mathcal{S}$-matrix. To transform to the `in-in' version, we will just need to insert a complete set of $|\rm{out}\rangle$ states. While we will see that this is indeed the case, the mechanism by which the soft and hard charges balance is slightly different. It will thus behoove us to spend a minute recasting the statement of the Ward identity~\eqref{eq:wardid} in terms of operators acting on $|\rm{in}\rangle$ and $|\rm{out}\rangle$. Here we will proceed by first following this route: deriving the `in-in' ward identity from the $\cal S$-matrix one. We will then show that this matches the `in-in' computation directly.

\subsubsection{${\cal S}$-matrix to `in-in' Ward Identity}
In the celestial literature we typically write the Ward identity with an explicit copy of the $\mathcal{S}$-matrix acting after/before $\mathcal{Q}^\pm$. As we noted in footnote~\ref{ft:inoutconv}, the notation in the celestial literature is somewhat different than what was introduced in section~\ref{sec:inin}, in that we would have written $\langle \mbox{out} |\mathcal{S}|\mbox{in}\rangle$ rather than  $\langle \mbox{out}|\mbox{in}\rangle$ as an $\mathcal{S}$-matrix element. The former is implicitly using $a,a^\dagger$ for the out state, while the latter uses $b,b^\dagger$, and the $\mathcal{S}$-matrix gives us that change of basis. Using the notation in section~\ref{sec:inin} we can write~\eqref{eq:wardid} as  
\begin{equation}\label{eq:wardab}
\langle \mbox{out}|\mathcal{Q}^+(b)|\mbox{in}\rangle = \langle \mbox{out}|\mathcal{Q}^-(a)|\mbox{in}\rangle 
\end{equation}
or, in terms of a particular $m\rightarrow n$ process
\be
0=\langle 0|b_{m+n}\cdots b_{m+1}(\mathcal{Q}^+(b)-\mathcal{Q}^-(a)) a^\dagger_{m}\cdots a^\dagger_1|0\rangle.
\ee
Applying~\eqref{eq:abbasis} and~\eqref{eq:abvac} tells us we can write this as
\be
0=\langle 0|a_{m+n}\cdots a_{m+1}\mathcal{S}(\mathcal{S}^\dagger \mathcal{Q}^+(a) \mathcal{S}-\mathcal{Q}^-(a)) a^\dagger_{m}\cdots a^\dagger_1|0\rangle
\ee
or, equivalently we can write this in terms of a commutator of the charge with the $\mathcal{S}$-matrix,
\be
0=\mathcal{Q}^+(a) \mathcal{S}-\mathcal{S}\mathcal{Q}^-(a)~~\Leftrightarrow~~0=\mathcal{Q}^+(b)-\mathcal{Q}^-(a),
\ee
at least when sandwiched between any Fock space state. The first expression is closer to the $0=\left[\mathcal{Q},\mathcal{S}\right]$ statement commonly made in the soft physics literature~\cite{Strominger:2017zoo} while the latter is how we would rewrite it given the notation from section~\ref{sec:inin}. Inserting a complete set of states results in
\be\badat{3}\label{eq:wardin}
\langle \mbox{in}'|(\mathcal{Q}^+(b)-\mathcal{Q}^-(a)) |\mbox{in}\rangle 
&=\sum_{out}\langle \mbox{in}'|\mbox{out}\rangle\langle \mbox{out}|(\mathcal{Q}^+(b)-\mathcal{Q}^-(a)) |\mbox{in}\rangle =0
\eadat\ee
where this quantity vanishes by virtue of~\eqref{eq:wardab}. 
Thus, the Ward identity for the `in-in' correlators follows simply from the one for $\mathcal{S}$-matrix elements.

The derivation as presented in section~\ref{sec:softWard} requires that $\langle \mbox{out}|a_\pm(\omega)^\dagger$ vanishes when we take $\omega$ soft, or that there could be additional contributions to the soft charge action $\mathcal{Q}_S^+$. While the earlier derivation is fine at tree level, these contributions would matter when we sum over all $|\rm{out}\rangle$ or when there are soft clouds. This is seemingly corroborated by Weinberg's statement that amplitudes without any soft gravitons vanish~\cite{Weinberg:1965nx}. However, in~\cite{Kapec:2017tkm} Strominger and collaborators showed that one can re-interpret the cancellation between real and virtual soft emissions -- a natural motivation to consider inclusive quantities -- precisely as a manifestation of asymptotic charge conservation in the $\mathcal{S}$-matrix. As such, the Ward identity indeed holds more generally ~\cite{Kapec:2017tkm,Gabai:2016kuf}. We make this comment since the $a_\pm(\omega)^\dagger$ terms in $\mathcal{Q}_S^+$ contribute to the `in-in' correlators, highlighting a fundamental difference in the two computations. In what follows, we will show that the Ward identity remains unchanged due to a matching modification to the action of $\mathcal{Q}_S^-$. 

We need to first emphasize one aspect of how~\eqref{eq:ward2} works out that we alluded to briefly at the end of section~\ref{sec:softWard}. The contributions from $\mathcal{Q}_S^+$ and $-\mathcal{Q}_S^-$ are equal and thereby double one another due to the form of the soft factor (shown for an additional outgoing graviton in~\eqref{eq:softfac}). For either an incoming or an outgoing soft graviton, it will couple to all of the external legs, however one can see from momentum conservation at the vertex that we get a relative sign from the propagator that gives the Weinberg pole depending on whether the soft graviton and the external leg are both out/in going or if one is incoming and one is outgoing. As such, there is an overall sign for the $\mathcal{Q}_S^-$ insertion. Meanwhile the $\mathcal{Q}_H^\pm$ only pick up contributions from the out/in state they are acting on. The factor of 1/2 one gets from averaging the $a_+$ and $a_-^\dagger$ contributions to $\mathcal{Q}_S^+$ and $\mathcal{Q}_S^-$ separately in~\eqref{eq:intNzz} exactly makes the soft and hard contributions match. 

\begin{figure}[htb]
    \centering
    \includegraphics[width=0.9\linewidth]{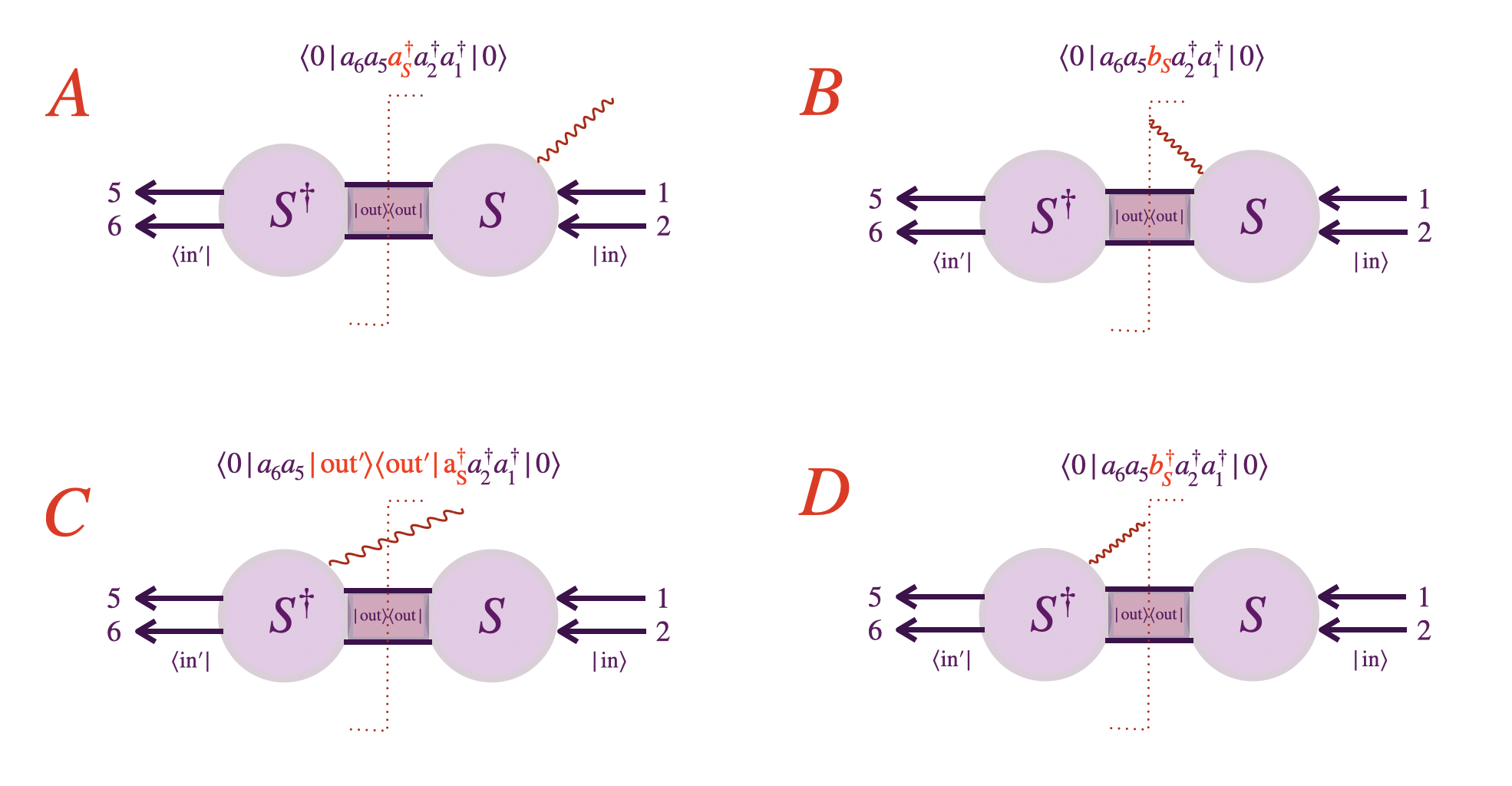}
    \caption{Relationship between crossing and type-I and II contributions to the soft factors. The red contributions to the correlators are due to the soft insertion. In diagram C, the state $|\rm{out}'\rangle$ contains an implicit sum over a soft particle.}
    \label{fig:ABCD}
\end{figure}

Computing the expectation value of a charge insertion in an `in-in' correlator presents some challenges that are not seen in the usual Ward identity derivation. For example, the expectation value of an additional incoming soft graviton vanishes in an `in-in' correlator since
\be
\langle 0| a_{m'}\cdots a_{1'} a_S^\dagger  a^\dagger_{m}\cdots a^\dagger_1 |0\rangle =0
\ee
and similarly for $a_S$. As such, we see that
\be
\langle \mbox{in}'|\mathcal{Q}_S^-(a_S,a_S^\dagger)|\mbox{in}\rangle =0 
\ee
so the one point function of this component of the soft charge trivially vanishes in this formalism. If we insert a sum over out states 
\be\label{eq:sum0}
\sum_{out}\langle \mbox{in}'|\mbox{out}\rangle\langle \mbox{out}|\mathcal{Q}_S^-|\mbox{in}\rangle =0
\ee
we might think that either $\langle{\rm{out}}|\mathcal{Q}_S^-$ or $\mathcal{Q}_S^-|{\rm{in}}\rangle$ trivially vanishes. However, the cancellation of terms leading to this is more subtle and we can see it as follows. Since $\mathcal{Q}_S^-$ contains a sum of a creation and annihilation operator, it is the sum of two terms represented by diagrams A and C 
in~\ref{fig:ABCD}. 
For concreteness let 
\be\badat{3}\label{eq:abcd}
A&:=\lim\limits_{\omega\rightarrow 0}\omega\langle \mbox{in}'|\mbox{out} \rangle\langle \mbox{out}|a_\pm^\dagger |\mbox{in}\rangle,~~&B&:= \lim\limits_{\omega\rightarrow 0}\omega\langle \mbox{in}'|\mbox{out} \rangle\langle \mbox{out}|b_\mp |\mbox{in}\rangle, \\
C&:= \lim\limits_{\omega\rightarrow 0}\omega \sum_{S} \langle \mbox{in}'|\mbox{out},S \rangle\langle \mbox{out},S|a_\pm^\dagger |\mbox{in}\rangle,~~
&D&:= \lim\limits_{\omega\rightarrow 0}\omega\langle \mbox{in}'|b_\pm^\dagger |\mbox{out} \rangle\langle \mbox{out}|\mbox{in}\rangle\\
\eadat\ee
where we are projecting onto a fixed $|\mbox{out}\rangle$ state, and are taking $|\rm{in}'\rangle\mapsto | \rm{in}\rangle$ in the end. Diagrams A, B and D are relatively straightforward to understand in the usual context. Diagram C is a little different because the soft particle actually crosses the cut. This means that a soft particle is present in the complete set of states we are summing over. Since crossing a soft graviton from an $a_\pm^\dagger$ insertion to a $b_{\mp}$ insertion introduces a sign we see that
\be
A=-B .
\ee
Meanwhile our complete set of $|\rm{out}\rangle$ will also include ones like figure $C$ and we expect $C=D$. At the same time, in the forward limit, $D$ should be none other than
\be\badat{3}
D&=\lim\limits_{\omega\rightarrow 0}\omega (\langle \mbox{out}|b_\pm |\mbox{in}\rangle)^* \langle \mbox{out}|\mbox{in}\rangle=\lim\limits_{\omega\rightarrow 0}\omega (S^{(0)\pm})^*\langle \mbox{in}' |\mbox{out} \rangle \langle \mbox{out}|\mbox{in}\rangle \\
&=\lim\limits_{\omega\rightarrow 0}\omega S^{(0)\mp}\langle \mbox{in}' |\mbox{out} \rangle \langle \mbox{out}|\mbox{in}\rangle=\lim\limits_{\omega\rightarrow 0}\omega \langle \mbox{in}' |\mbox{out} \rangle \langle \mbox{out}|b^\dagger_\mp|\mbox{in}\rangle=B 
\eadat\ee
so $D=B$. Note that the third equality follows from the complex conjugation of the polarization tensors and the last equality is just recognizing that the soft theorem could have also come from the insertion in $B$. Putting all of these together we see that
\be\label{eq:ac}
A=-C .
\ee
Finally, summing over $|\mbox{out}\rangle$ states gives~\eqref{eq:sum0}, even though we've organized things to cancel term by term by grouping contributions with an additional  outgoing soft graviton. 

As we mentioned, only diagrams $A,C$ were relevant for the charge one-point function vanishing above. The relation $B=D$ is important for another reason. Recall that~\eqref{eq:intNzz} tells us
\begin{equation}
{\cal N}_{zz}=-\frac{\kappa}{4 \pi(1+z \bar{z})^2} \lim _{\omega \rightarrow 0^{+}}\left[\omega b_{+}(\omega \hat{x})+\omega b_{-}(\omega \hat{x})^{\dagger}\right]
\end{equation}
which means the soft charge, written as
\be
\mathcal{Q}_S^+=-\frac{1}{16 \pi G} \int d^2 z \sqrt{\gamma} f\left[D_z^2 {\cal N}^{z z}+D_{\bar{z}}^2 {\cal N}^{\bar{z} \bar{z}}\right]
\ee
involves both $b_\pm$ and $b_\mp^\dagger$ terms.  In the usual $\mathcal{S}$-matrix ward identity we drop the $b^\dagger$ terms but in this derivation we cleverly insert two complete sets of states 
\be
\badat{3}
\langle \mbox{in}'|{\cal N}_{zz}|\mbox{in}\rangle &=\sum_{out,out'}\langle \mbox{in}'|\mbox{out}'\rangle\langle \mbox{out}'|{\cal N}_{zz}|\mbox{out}\rangle\langle \mbox{out}|\mbox{in}\rangle\\
&\propto \lim\limits_{\omega\rightarrow0}\sum_{out,out'}\langle \mbox{in}'|\mbox{out}'\rangle\langle \mbox{out}'|b_++b_-^\dagger|\mbox{out}\rangle\langle \mbox{out}|\mbox{in}\rangle\\
&\propto \lim\limits_{\omega\rightarrow0}\sum_{out}\langle \mbox{in}'|\mbox{out}\rangle\langle \mbox{out}|b_+ \mbox{in}\rangle+\langle \mbox{in}'|b_-^\dagger|\mbox{out}\rangle\langle \mbox{out}|\mbox{in}\rangle\\
\eadat\ee
and similarly for $\langle \mbox{out}|\mathcal{Q}_S^+|\mbox{in}\rangle$, i.e. terms of the form $D+B$. Therefore, while $\langle \mbox{in}'| \mathcal{Q}_S^-|\mbox{in}\rangle$ vanishes, $\langle \mbox{in}'|\mathcal{Q}_S^+|\mbox{in}\rangle$ doubles and so $\langle \mbox{in}'| \mathcal{Q}_S^+-\mathcal{Q}_S^-|\mbox{in}\rangle$ stays the same! In particular, the additional soft modes in the out state resolve the tension between the fact that $\mathcal{Q}_S^\pm$ have different expectation values in `in-in' correlators and the fact that in $\mathcal{S}$-matrix elements $a^\dagger_\pm$ crosses to the $b_\mp$ insertion which has the same value (up to a sign) in the soft limit. 

Meanwhile, $\mathcal{Q}_H^\pm$ does not change the particle number (and also is insensitive to soft gravitons) and is thereby unchanged. We see that the naive Ward identity~\eqref{eq:wardin} indeed holds, albeit with a bit of a reshuffling of how the soft terms contribute as compared to the $\mathcal{S}$-matrix version~\eqref{eq:wardab}. We are now in a position to use this Ward identity to rewrite $\mathcal{Q}_S^+$ in terms of the matter fields using `in-in' correlators. But before we do that, we will establish some notation to simplify the expressions that we get. The hard charge is diagonalized on either $|\rm{in}\rangle$ or $|\rm{out}\rangle$, depending on whether it is $\mathcal{Q}_H^{\pm}$. Therefore, we will write 
\begin{equation}
\langle \mbox{out}| \mathcal{Q}_{H}^+(b)\equiv \mathbb{Q}_H^+\langle \mbox{out}|, \ \ \mathcal{Q}_H^-(a)|\mbox{in}\rangle\equiv\mathbb{Q}_H^-|\mbox{in}\rangle
\end{equation}
where the $\mathbb{Q}_H^\pm$ are the respective eigenvalues of the hard charge. It is also important to note that if one adds a soft particle to the out state, the action of the hard charge on that state will be the same since the additional contribution to the eigenvalue will be proportional to the energy of the soft particle which is taken to 0. We will also use the notation that 
\begin{eqnarray}\label{eq:aneconetwo}
\mathbb{E}_1(\vec{z}_S)  & \equiv & \langle \mbox{in}'|\mathcal{Q}_H^+(b)|\mbox{in}\rangle \equiv  \sum_{\rm{out}}\mathbb{Q}_H^+\langle \mbox{in}'|\mbox{out}\rangle\langle \mbox{out}|\mbox{in}\rangle \cr
\mathbb{E}_2(\vec{z}_{S,1};\vec{z}_{S,2}) & \equiv & \langle \mbox{in}'|\mathcal{Q}_{H,1}^+(b)\mathcal{Q}_{H,2}^+(b)|\mbox{in}\rangle 
\end{eqnarray}
where $\mathbb{E}_1,\mathbb{E}_2$ are the one-point and two-point functions of the ANEC that we defined earlier. Note that there are no superscripts on $\mathbb{E}_1,\mathbb{E}_2$ since this statement is expected to be true at all orders and not just for connected contributions. Putting this together we have that for massless scattering
\be\badat{3}\label{eq:QS+}
\langle \mbox{in}'| \mathcal{Q}_S^+|\mbox{in}\rangle&=\langle \mbox{in}'| \mathcal{Q}_S^+(b)-\mathcal{Q}_S^-(a)|\mbox{in}\rangle=-\langle \mbox{in}'| \mathcal{Q}_H^+(b)-\mathcal{Q}_H^-(a)|\mbox{in}\rangle\\
&=-\mathbb{E}_1(\vec{z}_S) + 4\pi G \sum_{i\in in} \omega_i {\delta(\vec{z}_{Si})}/{\sqrt{\gamma}} 
\eadat\ee
in the forward limit $|\rm{in}'\rangle \mapsto |\rm{in}\rangle$ for the choice\footnote{This particular choice localizes the sphere integral so that it does not come along for the ride.} $f(z,\bz)=\delta(\vec{z}-\vec{z}_S)$. 

We notice that the first term is just the expectation of the ANEC operator, which has been computed in~\cite{Hofman:2008ar} for various CFTs,~\cite{Belitsky2014b} in ${\cal N}=4$ SYM, and~\cite{Herrmann:2024yai} for gravity. Moreover, we can see that because of the form of the soft factor, the derivatives hitting ${\cal N}_{AB}$ will be such that it localizes to each external leg that the soft graviton attaches to, as in equation~\eqref{eq:local}. Whenever it attaches to $|\rm{out}\rangle$ we get something proportional to an ANEC 1-point function, whereas whenever it attaches to $|\rm{in}\rangle$, we pick up the (known) energy of said state. If we chose to dress $|\rm{in}\rangle$, as in~\cite{Choi:2017ylo}, we could remove this term as well.

Alternatively if we choose $z_S$ so that we are away from any incoming particles, we can also drop the second term in the last line of~\eqref{eq:QS+}. This is different from the sum over $|\rm{out}\rangle$, which are integrated over locations. In particular the delta function from the soft factor that localizes $z_S$ to $z_i$ for $i\in out$, a particle in the sum over $|\rm{out}\rangle$, will have an integral over $z_i$ that localizes to the soft charge/ANEC location. This is precisely why the phase space integrals for an ANEC insertion are seemingly simpler than for a memory insertion. As soon as we smear the soft charge we should be careful not to drop the contributions from $|\rm{in}\rangle$. We will return to this when we discuss memory correlators in section~\ref{sec:memmem}.

\subsubsection{From `in-in' Insertions to Ward Identity}
While~\eqref{eq:QS+} was derived starting from the Ward identity for the ${\cal S}$-matrix, we can also check it directly from the `in-in' correlators. As mentioned previously, since the in-in computation has infinitely many terms corresponding to the number of particles in $|\rm{out}\rangle$, here we will explicitly verify the conclusions from the above computation to lowest order i.e fewest particles in $|\rm{out}\rangle$. It is also important to note that we work with the restricted kinematics in~\cite{Herrmann:2024yai} and that for our purposes, it is enough to consider the connected contributions to the amplitude.

The definitions of the relevant operators were defined in section~\ref{sec:celenergy}. Note that in~\cite{Herrmann:2024yai} they found expressions for insertions of detector operators and not the soft charges that we are discussing here. Therefore, we would like to insert a soft charge $\mathcal{Q}_S$ and then show that it is related to the one-point function of the detector operator we defined above. The insertion of a soft charge effectively adds a soft particle (graviton, in this case) to the external leg of each half of the diagram (i.e one scattering process). In both cases (up to an overall sign), the one point function will look like 
\begin{eqnarray}
\langle \mbox{in}|\mathcal{Q}_S^+(\vec{z}_S)|\mbox{in}\rangle & = & \int d^4 q \sum_{i=1}^n\frac{4\pi G}{\sqrt{\gamma}}\omega_i \delta(\vec{z}_{Si})\left|\mathcal{M}_{n\rightarrow 1}^{(0)}(\{p_n\},q)\right|^2\delta^{(4)}(q-P)\cr
& - &\int d^4 q \frac{4\pi G}{\sqrt{\gamma}}\omega \delta(\vec{z}_{S}-\vec{w})\left|\mathcal{M}_{n\rightarrow 1}^{(0)}(\{p_n\},q)\right|^2\delta^{(4)}(q-P).
\end{eqnarray}
We have used the form of the soft theorem here for the $n\rightarrow 1 $ amplitude and the soft momenta is labeled by $\vec{z}_S$. It is important to note that the right hand side of this equation has the full soft factor which allows the soft leg to attach to an incoming or outgoing leg while the left hand side of this equation has only the future part of the soft charge. This is because, as we showed earlier, the `in-in' one point function of $\mathcal{Q}_S^-$ vanishes. We can recognize that the second term above is the ANEC one-point function. Therefore we see that 
\begin{equation}
\langle \mbox{in}|\mathcal{Q}_S^+(\vec{z}_S)|\mbox{in}\rangle  =  \frac{4\pi G}{\sqrt{\gamma}}\sum_{i=1}^n\omega_i \delta(\vec{z}_{Si}) \left|\mathcal{M}_{n\rightarrow 1}^{(0)}(p_i,P)\right|^2-\mathbb{E}_1^{(0)}(\vec{z}_S).
\end{equation}
which matches the derivation in the previous section. This relatively simple computation illustrates that the Ward identity relation can be derived in the context of the operators defined more explicitly in~\cite{Herrmann:2024yai}.

\subsection{Soft Charge Correlators}\label{sec:QSQS}
We saw that the one point function of the soft charge could be related to the one point function of the ANEC in a way that was consistent with the standard supertranslation Ward identity. In much of the literature regarding energy-energy correlators, one is primarily interested in higher point functions of the ANEC~\cite{Henn:2019gkr,Belitsky:2013ofa,Dixon:2018qgp,three_chen_2020,Komiske:2022enw,spinning_chen_2022,Herrmann:2024yai,Chicherin:2024ifn,Chen:2019bpb,Ma:2025qtx,Yan:2022cye,Yang:2024gcn,Yang:2022tgm,Chen:2022swd}. Therefore, in this section we aim to understand how higher point correlators of the soft charges can be written in terms of ANEC correlators, focusing on the two point functions for concreteness. We will find that unlike what we might have expected from a naive application of the celestial amplitude OPEs~\cite{Himwich:2020rro,Pasterski:2021dqe}, the connected correlators can be nonzero. As in the previous section we will proceed in two steps: starting directly from the $\cal S$-matrix Ward identity and then checking against the `in-in' correlators.

\subsubsection{$\cal S$-matrix Ward Identity to Two-Point Functions}
We can use the Ward identity from the previous section to determine the form of the $\mathcal{Q}_S^+$ two point function, as follows. First, we note that we expect the charges to obey a representation of the BMS algebra, and since supertranslations commute, so should two copies of $\mathcal{Q}^+$ with each other (and similarly for $\mathcal{Q}^-$). Moreover one can check from the explicit form of the hard and soft charges that $[\mathcal{Q}^\pm_S,\mathcal{Q}^\pm_H]=0$ (see e.g.~\cite{Donnay:2021wrk}). Still, we need to be careful about how to order the $\mathcal{Q}^+$ and $\mathcal{Q}^-$ terms, as we see that these could involve more general time folds~\cite{Caron-Huot:2023vxl} than the ones set up in section~\ref{sec:inin} above.

The crux of the computation for the soft two-point function will come from the following statement
\be\badat{3}\label{eq:wardin2}
&\langle \mbox{in}'|\mathcal{Q}_{S,1}^+(b)(\mathcal{Q}_2^+(b)-\mathcal{Q}_2^-(a)) |\mbox{in}\rangle \\ 
&=\sum_{out,out'}\langle \mbox{in}'|\mbox{out}'\rangle\langle \mbox{out}'|\mathcal{Q}_{S,1}^+(b)|\mbox{out}\rangle\langle \mbox{out}|(\mathcal{Q}_2^+(b)-\mathcal{Q}_2^-(a)) |\mbox{in}\rangle =0
\eadat\ee
which uses the Ward identity conveniently in the second factor upon introducing a complete set of states. Grouping the soft and hard pieces equivalently gives
\be\badat{3}\label{eq:ward2exp}
&\langle \mbox{in}'|\mathcal{Q}_{S,1}^+(b)\mathcal{Q}_{S,2}^+(b) |\mbox{in}\rangle \\
&=-\langle \mbox{in}'|\mathcal{Q}_{S,1}^+(b)\mathcal{Q}_{H,2}^+(b) |\mbox{in}\rangle+\langle \mbox{in}'|\mathcal{Q}_{S,1}^+(b)\mathcal{Q}_{S,2}^-(a) |\mbox{in}\rangle+\langle \mbox{in}'|\mathcal{Q}_{S,1}^+(b)\mathcal{Q}_{H,2}^-(a) |\mbox{in}\rangle.
\eadat\ee
The left hand side of this equation is what we identify as the soft charge two point function. In order to compute it, we need to know the three terms on the right hand side. 

\paragraph{First term $\langle \mbox{in}'|\mathcal{Q}_{S,1}^+(b)\mathcal{Q}_{H,2}^+(b) |\mbox{in}\rangle$:} We can use the Ward identity to write this as 
\begin{eqnarray}
\langle \mbox{in}'|\mathcal{Q}_{S,1}^+(b)\mathcal{Q}_{H,2}^+(b) |\mbox{in}\rangle & = & -\langle \mbox{in}'|\mathcal{Q}_{H,1}^+(b)\mathcal{Q}_{H,2}^+(b) |\mbox{in}\rangle + \langle \mbox{in}'|\mathcal{Q}_{S,1}^-(a)\mathcal{Q}_{H,2}^+(b) |\mbox{in}\rangle\cr
& + & \langle \mbox{in}'|\mathcal{Q}_{H,1}^-(a)\mathcal{Q}_{H,2}^+(b) |\mbox{in}\rangle.
\end{eqnarray}
For the second term on the right hand side, we can insert a complete set of states which will make it proportional to $\langle \mbox{in}'|\mathcal{Q}_{S,1}^-(a)|\mbox{out}\rangle$. This vanishes since it adds a soft particle to $|\rm{in}\rangle$ resulting in a zero overlap. Therefore, there are only two contributions to this which we can write concisely as
\begin{equation}
\langle \mbox{in}'|\mathcal{Q}_{S,1}^+(b)\mathcal{Q}_{H,2}^+(b) |\mbox{in}\rangle  =  -\mathbb{E}_2(\vec{z}_{S,1};\vec{z}_{S,2})  +\mathbb{Q}_{H,1}^-\mathbb{E}_1(\vec{z}_{S,2}).
\end{equation}
Therefore, this first term can be written concisely in terms of the ANEC one and two point functions as we defined in~\eqref{eq:aneconetwo}.

\paragraph{Second term $\langle \mbox{in}'|\mathcal{Q}_{S,1}^+(b)\mathcal{Q}_{S,2}^-(a) |\mbox{in}\rangle$:} We first insert a complete set of states
\begin{equation}
\langle \mbox{in}'|\mathcal{Q}_{S,1}^+(b)\mathcal{Q}_{S,2}^-(a) |\mbox{in}\rangle = \sum_{out} \langle \mbox{in}'|\mathcal{Q}_{S,1}^+(b)|\mbox{out}\rangle\langle \mbox{out}|\mathcal{Q}_{S,2}^-(a) |\mbox{in}\rangle .
\end{equation}
Next, we use the Ward identity on the first factor to write this as three terms
\begin{eqnarray}
\langle \mbox{in}'|\mathcal{Q}_{S,1}^+(b)\mathcal{Q}_{S,2}^-(a) |\mbox{in}\rangle & = & - \langle \mbox{in}'|\mathcal{Q}_{H,1}^+(b)\mathcal{Q}_{S,2}^-(a) |\mbox{in}\rangle  +  \langle \mbox{in}'|\mathcal{Q}_{S,1}^-(a)\mathcal{Q}_{S,2}^-(a) |\mbox{in}\rangle\cr
& + &\sum_{out} \langle \mbox{in}'|\mathcal{Q}_{H,1}^-(a)|\mbox{out}\rangle\langle \mbox{out}|\mathcal{Q}_{S,2}^-(a) |\mbox{in}\rangle  
\end{eqnarray}
where we have collapsed the sum over out states in the first two terms. Finally, we can use the fact that $\mathcal{Q}_{H,1}^-(a)$ is diagonalized on $|\rm{in}\rangle $ to show that the third term vanishes since it is the one point function of $\mathcal{Q}_{S,2}^-(a)$
\begin{equation}
\sum_{out} \langle \mbox{in}'|\mathcal{Q}_{H,1}^-(a)|\mbox{out}\rangle\langle \mbox{out}|\mathcal{Q}_{S,2}^-(a) |\mbox{in}\rangle  = \mathbb{ Q}_{H,1}^-  \cancel{\langle \mbox{in}'|\mathcal{Q}_{S,2}^-(a) |\mbox{in}\rangle}
\end{equation}
which means that this two point function only has two terms remaining
\begin{equation}
\langle \mbox{in}'|\mathcal{Q}_{S,1}^+(b)\mathcal{Q}_{S,2}^-(a) |\mbox{in}\rangle = -\langle \mbox{in}'|\mathcal{Q}_{H,1}^+(b)\mathcal{Q}_{S,2}^-(a) |\mbox{in}\rangle
+\langle \mbox{in}'|\mathcal{Q}_{S,1}^-(a)\mathcal{Q}_{S,2}^-(a) |\mbox{in}\rangle .  
\end{equation}
The first term on the right hand side vanishes because if we insert a complete set of states then the sum will vanish term by term as per~\eqref{eq:sum0}. Therefore 
\begin{equation}
\langle \mbox{in}'|\mathcal{Q}_{S,1}^+(b)\mathcal{Q}_{S,2}^-(a) |\mbox{in}\rangle = \langle \mbox{in}'|\mathcal{Q}_{S,1}^-(a)\mathcal{Q}_{S,2}^-(a) |\mbox{in}\rangle 
\end{equation}
and this term is quite simple.

\paragraph{Third term $\langle \mbox{in}'|\mathcal{Q}_{S,1}^+(b)\mathcal{Q}_{H,2}^-(a) |\mbox{in}\rangle$:} Noting that $\mathcal{Q}^-_H(a)$ is diagonalized on the in state allows us to write this as
\begin{equation}
\langle \mbox{in}'|\mathcal{Q}_{S,1}^+(b)\mathcal{Q}_{H,2}^-(a) |\mbox{in}\rangle =\mathbb{Q}_{H,2}^-\langle \mbox{in}'|\mathcal{Q}_{S,1}^+(b)|\mbox{in}\rangle.
\end{equation}
Next, we can use the Ward identity to write the one point function of the soft charge as 
\begin{equation}
\langle \mbox{in}'|\mathcal{Q}_{S,1}^+(b)\mathcal{Q}_{H,2}^-(a) |\mbox{in}\rangle=\mathbb{Q}_{H,2}^-[-\langle \mbox{in}'|\mathcal{Q}_{H,1}^+(b)|\mbox{in}\rangle+\cancel{\langle \mbox{in}'|\mathcal{Q}_{S,1}^-(a)|\mbox{in}\rangle}+\langle \mbox{in}'|\mathcal{Q}_{H,1}^-(a)|\mbox{in}\rangle]
\end{equation}
The second term vanishes since $\mathcal{Q}_{S,1}^-(a)$ will add a soft graviton to $|\rm{in}\rangle$ which will have zero overlap with itself, yielding a vanishing inner product. Therefore, an insertion of one component of the soft charge and one component of the hard charge gives only two terms
\begin{equation}
\langle \mbox{in}'|\mathcal{Q}_{S,1}^+(b)\mathcal{Q}_{H,2}^-(a) |\mbox{in}\rangle=-\mathbb{Q}_{H,2}^-\mathbb{E}_1(\vec{z}_{S,1})+\mathbb{Q}_{H,2}^-\mathbb{ Q}_{H,1}^-.
\end{equation}
This third time can thereby be written in terms of an ANEC one-point function and eigenvalues of the hard charge on $|\rm{in}\rangle$.

We are now in a position to combine all three terms together to write the soft charge two point function. We see that it is given by 
\begin{eqnarray}\label{eq:plplward}
\langle \mbox{in}'|\mathcal{Q}_{S,1}^+(b)\mathcal{Q}_{S,2}^+(b) |\mbox{in}\rangle &= & \mathbb{E}_2(\vec{z}_{S,1};\vec{z}_{S,2})  -\mathbb{Q}_{H,1}^-\mathbb{E}_1(\vec{z}_{S,2})-\mathbb{Q}_{H,2}^-\mathbb{E}_1(\vec{z}_{S,1})\cr
& + & \langle \mbox{in}'|\mathcal{Q}_{S,1}^-(a)\mathcal{Q}_{S,2}^-(a) |\mbox{in}\rangle +\mathbb{Q}_{H,2}^-\mathbb{ Q}_{H,1}^-.
\end{eqnarray}
This formula is quite nice because we see that the two point function of the soft charges can be computed using the one- and two-point functions of the ANECs. It is also useful to figure out the connected component of this two point function which means we subtract off the disconnected diagrams. In this case it amounts to 
\begin{equation}
\langle \mbox{in}'|\mathcal{Q}_{S,1}^+(b)\mathcal{Q}_{S,2}^+(b) |\mbox{in}\rangle_c = \langle \mbox{in}'|\mathcal{Q}_{S,1}^+(b)\mathcal{Q}_{S,2}^+(b) |\mbox{in}\rangle-\langle \mbox{in}'|\mathcal{Q}_{S,1}^+(b)|\mbox{in}\rangle \langle \mbox{in}'|\mathcal{Q}_{S,2}^+(b) |\mbox{in}\rangle.
\end{equation}
We can compute this by first noting that the Ward identity tells us 
\begin{equation}
\langle \mbox{in}'|\mathcal{Q}_{S}^+(b) |\mbox{in}\rangle \equiv -\mathbb{E}_1 + \mathbb{Q}_H^-
\end{equation}
where we have used the fact that $\langle \mbox{in}'|\mathcal{Q}_{S}^-(a) |\mbox{in}\rangle = 0$, the action of $\mathcal{Q}_H^-(a)$ on in states and the definition of the ANEC one-point function. Then we can substitute that in for the disconnected component and substitute the full two point function from above 
\begin{equation}\label{eq:2ptconn}
\langle \mbox{in}'|\mathcal{Q}_{S,1}^+(b)\mathcal{Q}_{S,2}^+(b) |\mbox{in}\rangle_c = \mathbb{E}_{2,c}(\vec{z}_{S,1};\vec{z}_{S,2}) + \langle \mbox{in}'|\mathcal{Q}_{S,1}^-(a)\mathcal{Q}_{S,2}^-(a) |\mbox{in}\rangle  
\end{equation}
where we have denoted the connected part of the ANEC two point function to be 
\begin{equation}
\mathbb{E}_{2,c}(\vec{z}_{S,1};\vec{z}_{S,2})\equiv \mathbb{E}_2(\vec{z}_{S,1};\vec{z}_{S,2})-\mathbb{E}_1(\vec{z}_{S,1}) \mathbb{E}_1(\vec{z}_{S,2}).
\end{equation}
The second term, involving two incoming soft charges, is just a contact term, and will vanish if we pick the smearing functions $f_{1}$ and $f_2$ with disjoint support.
These terms will only matter if we smear over the ANEC locations, which will show up in the next section. Note that in computing the connected correlators all of the $\mathbb{Q}^-_H$ terms have dropped. While for the one point functions we could remove these terms by dressing the in states, the connected two point functions simplify without needing to do so.

\subsubsection{Soft Charge Two-Point Function to Vertex Insertions}

As we did for the one point function, we can also compute the two point function of soft charges and relate them to the vertex functions that we defined earlier. We will utilize the fact that for the leading soft limit in gravity, there is no issue with compounding soft factors~\cite{Weinberg:1965nx}. In this case, we will need to consider the second type of diagram in Figure~\ref{fig:one_point} where there are two out particles in the out state because that is the easiest order to make contact with the ANEC two point function written in terms of vertex functions. Once again, we will be using restricted kinematics as in~\cite{Herrmann:2024yai} and matching the computation to the leading order of the connected amplitude.

As in the second diagram of Figure~\ref{fig:one_point} we will denote the outgoing momenta as $q_1,q_2$ and incoming momenta as $p_1,\cdots,p_n$. Two insertions of the soft charge will then result in the product of two soft factors as follows
\begin{eqnarray}
\langle \mbox{in}|\mathcal{Q}^+_{S_1}(\vec{z}_{S_1})\mathcal{Q}^+_{S_2}(\vec{z}_{S_2})|\mbox{in}\rangle & = & \int d^4 q_1 d^4 q_2 \frac{(4\pi G)^2}{\gamma}\left|\mathcal{M}_{n\rightarrow 2}^{(0)}\left(\{p_n\},q_1,q_2\right)\right|^2\delta^{(4)}(q_1+q_2-P)\cr
& \times & \left[\sum_{i=1}^n\omega_i\delta(\vec{z}_{S,1}-\vec{z}_i)-\sum_{j=1}^2\omega_j\delta(\vec{z}_{S,1}-\vec{w}_j)\right]\cr
& \times & \left[\sum_{k=1}^n\omega_i\delta(\vec{z}_{S,2}-\vec{z}_k)-\sum_{\ell=1}^2\omega_j\delta(\vec{z}_{S,2}-\vec{w}_\ell)\right].
\end{eqnarray}
This will give rise to four terms when we expand each of the soft factors
\begin{eqnarray}
\langle \mbox{in}|\mathcal{Q}^+_{S_1}(\vec{z}_{S_1})\mathcal{Q}^+_{S_2}(\vec{z}_{S_2})|\mbox{in}\rangle & = & \int d^4 q_1 d^4 q_2 \frac{(4\pi G)^2}{\gamma}\left|\mathcal{M}_{n\rightarrow 2}^{(0)}\left(\{p_n\},q_1,q_2\right)\right|^2\delta^{(4)}(q_1+q_2-P)\cr
& \times & \sum_{i,j=1}^n\omega_i\omega_j\delta(\vec{z}_{S,1}-\vec{z}_i)\delta(\vec{z}_{S,2}-\vec{z}_j)\cr
& - & \int d^4 q_1 d^4 q_2 \frac{(4\pi G)^2}{\gamma}\left|\mathcal{M}_{n\rightarrow 2}^{(0)}\left(\{p_n\},q_1,q_2\right)\right|^2\delta^{(4)}(q_1+q_2-P)\cr
& \times & \sum_{i=1}^n\sum_{j=1}^2\omega_i\omega_j\delta(\vec{z}_{S,1}-\vec{w}_j)\delta(\vec{z}_{S,2}-\vec{z}_i)\cr
& - & \int d^4 q_1 d^4 q_2 \frac{(4\pi G)^2}{\gamma}\left|\mathcal{M}_{n\rightarrow 2}^{(0)}\left(\{p_n\},q_1,q_2\right)\right|^2\delta^{(4)}(q_1+q_2-P)\cr
& \times & \sum_{i=1}^n\sum_{j=1}^2\omega_i\omega_j\delta(\vec{z}_{S,1}-\vec{z}_i)\delta(\vec{z}_{S,2}-\vec{w}_j)\cr
& + & \int d^4 q_1 d^4 q_2 \frac{(4\pi G)^2}{\gamma}\left|\mathcal{M}_{n\rightarrow 2}^{(0)}\left(\{p_n\},q_1,q_2\right)\right|^2\delta^{(4)}(q_1+q_2-P)\cr
& \times & \sum_{i,j=1}^2\omega_i\omega_j\delta(\vec{z}_{S,1}-\vec{w}_i)\delta(\vec{z}_{S,2}-\vec{w}_j).
\end{eqnarray}
We can look at each of these terms separately. The first term can be identified as the product of the eigenvalues of the hard charge on the in-states since it is just the product of those two delta functions. Therefore we can write it as $\mathbb{Q}_{H,1}^-\mathbb{Q}_{H,2}^-$. Skipping the middle two terms for now, we can look at the last term. We see that when $i=j$, it will vanish since we are assuming that the two soft particles are not coincident. Therefore, we only need the case when $i\neq j$. If we compare to~\eqref{eqn:leadingtwo} we see that this is precisely the ANEC two point function $\mathbb{E}_2^{(0)}$. Now we can look at the middle two terms by writing them as
\begin{eqnarray}
\mbox{Diff} & = & \langle \mbox{in}|\mathcal{Q}^+_{S_1}(\vec{z}_{S_1})\mathcal{Q}^+_{S_2}(\vec{z}_{S_2})|\mbox{in}\rangle - \mathbb{Q}_{H,1}^-\mathbb{Q}_{H,2}^--\mathbb{E}_2^{(0)}\cr
& = & -\sum_{i=1}^n\omega_i\delta(\vec{z}_{S,2}-\vec{z}_i)\int d^4 q_1 d^4 q_2 \frac{(4\pi G)^2}{\gamma}\left|\mathcal{M}_{n\rightarrow 2}^{(0)}\left(\{p_n\},q_1,q_2\right)\right|^2\cr
& \times & \sum_{j=1}^2\omega_j\delta(\vec{z}_{S,1}-\vec{w}_j)\delta^{(4)}(q_1+q_2-P)\cr
& - & \sum_{i=1}^n\omega_i\delta(\vec{z}_{S,1}-\vec{z}_i)\int d^4 q_1 d^4 q_2 \frac{(4\pi G)^2}{\gamma}\left|\mathcal{M}_{n\rightarrow 2}^{(0)}\left(\{p_n\},q_1,q_2\right)\right|^2\cr
& \times & \sum_{j=1}^2\omega_j\delta(\vec{z}_{S,2}-\vec{w}_j)\delta^{(4)}(q_1+q_2-P)\cr
& = & -\mathbb{Q}_{H,2}^-\mathbb{E}_1(\vec{z}_{S,1})-\mathbb{Q}_{H,1}^-\mathbb{E}_1(\vec{z}_{S,2}).
\end{eqnarray}
Putting all of this together we see that 
\begin{equation}
\langle \mbox{in}|\mathcal{Q}^+_{S_1}(\vec{z}_{S_1})\mathcal{Q}^+_{S_2}(\vec{z}_{S_2})|\mbox{in}\rangle = \mathbb{Q}_{H,1}^-\mathbb{Q}_{H,2}^-+\mathbb{E}_2^{(0)}-\mathbb{Q}_{H,2}^-\mathbb{E}_1(\vec{z}_{S,1})-\mathbb{Q}_{H,1}^-\mathbb{E}_1(\vec{z}_{S,2}).
\end{equation}
Comparing to~\eqref{eq:plplward} we see that this is the same expression without the contact term between the soft charges. This is to be expected because our assumption was that we chose the supports to be disjoint i.e. we only computed the connected components of the amplitudes. We therefore conclude that the soft two point function can be written in terms of the ANEC one-point and two-point functions, which is consistent with what we obtained before. We note that we expect this to hold at all orders even though we have computed only the leading order contribution in this case.

Nothing stops us from applying similar techniques to higher point correlators to similarly reduce them to ANEC expressions, though we expect the intermediate manipulations to get messier the more operators we add. Note that barring contact terms, it would appear that at one and two point the soft charge and ANEC correlators take the same form for dressed in states. This already suggests two differences from the soft mode correlators implied by the celestial OPE~\cite{Himwich:2020rro,Pasterski:2021dqe}: first, the soft-charge soft-charge two point functions are non-zero. Second, it would seem that there is no reason for the correlators to stop at two points. This would seem to suggest an interesting modification to the soft effective action~\cite{Kapec:2021eug} we if tried to model it as generating these terms (see~\cite{Agarwal:2021ais} for interesting work in the case of non-abelian gauge theories). We will now turn to the memory modes that the results in~\cite{Himwich:2020rro,Pasterski:2021dqe} were written in, so we can see this more explicitly.

\section{Memory Correlators}\label{sec:memmem}
In the previous section we demonstrated that the Ward identity for supertranslations holds for `in-in' correlators which implies that the soft charges have a non-trivial connected two point function. In this section, we will use the celestial diamond framework of~\cite{Pasterski:2021fjn,Pasterski:2021dqe} to lift our observations about $\mathcal{Q}_S^+$ correlators to statements about the memory effect.

Recall from equation~\eqref{mem}, that the soft modes appearing in $\mathcal{Q}_S$ were related to DC shifts in the metric
\be
\mathcal{N}_{zz}=\int du N_{zz}= C_{zz}|_{u=-\infty}^{u=\infty}\equiv \Delta C_{zz}.
\ee
The soft theorems imply that these long range tails take a universal form, determined by the scattering data. Indeed the pattern of relations between soft theorems, Ward identities and memory effects (the so-called infrared triangle~\cite{Strominger:2014pwa,Pasterski:2015tva,Pasterski:2015zua}) generalizes to other gauge theories and asymptotic symmetries. 
\begin{figure}[thb!]
    \centering
    \includegraphics[width=0.9\linewidth]{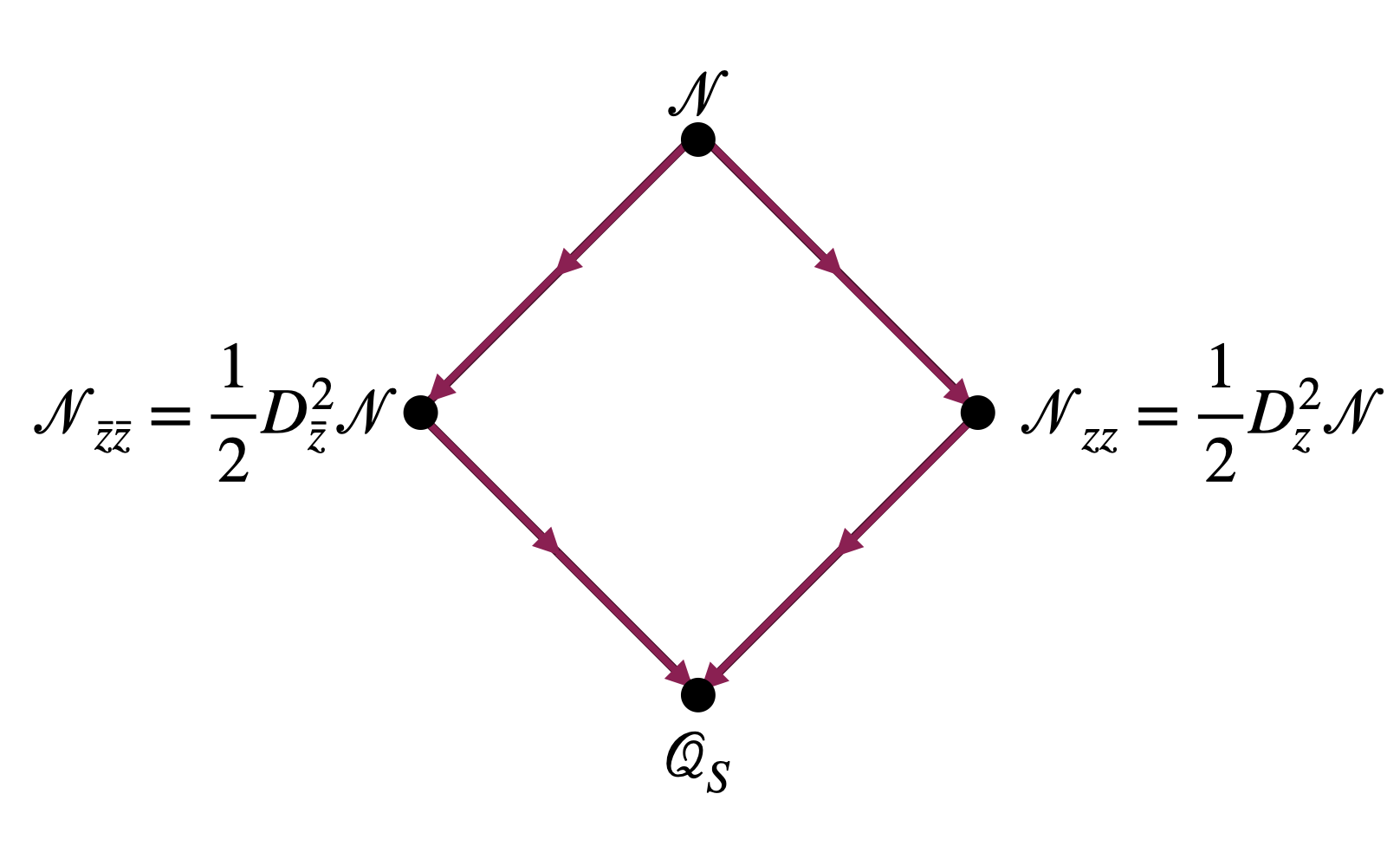}
    \caption{
Memory celestial diamond~\cite{Pasterski:2021dqe}. The descendant of the radiative memory modes $\mathcal{N}_{zz}$ and $\mathcal{N}_{\bar{z}\bar{z}}$ is the soft charge since it is given by two derivatives, notated by the two arrows, acting on the memory modes. }
    \label{fig:CN}
\end{figure}

These differential relations between the memory modes and the soft charge are captured by the celestial diamonds~\cite{Pasterski:2021fjn,Pasterski:2021dqe} as illustrated in figure~\ref{fig:CN}.\footnote{The memory diamond discussed here comes with a partner diamond called the goldstone diamond. The corners of both diamonds are related by a symplectic pairing.} The statement here is that the leading soft mode is secretly a scalar degree of freedom with the memory modes of opposite helicities, $\mathcal{N}_{zz}$ and $\mathcal{N}_{\bar{z}\bar{z}}$, descending to the same soft charge $\mathcal{Q}_S$. Note that we can also lift the memory modes to a $\Delta=-1,J=0$ celestial scalar which is labeled by $\mathcal{N}$. This is unnecessary except insofar as it puts the two-point functions in our infrared effective theory into a symmetric form
\begin{equation}\label{eq:softcorr}
\left\langle\mathcal{N}\left(w_i, \bar{w}_i\right) \mathcal{N}\left(w_j, \bar{w}_j\right)\right\rangle=\frac{1}{4} k_{\mathcal{N} \mathcal{N}}\left|w_{i j}\right|^2 \log \left|w_{i j}\right|^2
\end{equation}
Starting from the celestial OPEs~\cite{Himwich:2020rro}, it was conjectured in~\cite{Pasterski:2021dqe} that $k_{\mathcal{N}\mathcal{N}}=0$. In the gaussian model assuming $k_{\mathcal{N} \mathcal{N}}=0$ would then seem to match both the infrared divergent parts of amplitudes and the measurement of the soft charge. The only non-zero levels came from correlators involving the conjugate Goldstone mode $\mathcal{C}$ that capture Wilson line dressings~\cite{Himwich:2020rro,Arkani-Hamed:2020gyp}, and for which $k_{\mathcal{C}\mathcal{C}}$ is infrared divergent. 

There are two things apparent from this discussion: first, drawing conclusions about correlation functions from amplitudes is a bit misleading since they are not expectation values in the standard sense. By contrast the `in-in' correlators are measuring the memory's expectation for a given initial state. In our computations above we see that the soft charge two-point function would be expected to be non-zero if we had started with the assumption that $k_{\mathcal{N}\mathcal{N}}=0$ since any number of derivatives on that will still be 0. This seems contrary to~\eqref{eq:softcorr}, though we should note that the form of~\eqref{eq:softcorr} is determined by the dimensions of the operators and assuming that we are in vacuum, whereas the in state in our `in-in' correlator is expected to break some of the symmetries. Second, while the toy model above is Gaussian, we would not seem to expect such a trivialization of the ANEC correlators.

While we expect the form of the correlators to be modified, the celestial diamond picture is still relevant in the following way: the two helicity soft insertions are still related by a shadow transform (i.e. this is a statement that follows from the form of the soft factor and not some toy model) and this implies that we should be able to lift the soft charge to the radiative modes. Reminding ourselves that 
\be
\mathcal{Q}_S^+=-\frac{\sqrt{\gamma}}{8 \pi G} D_z^2 {\cal N}^{z z}
\ee
where we have already used the fact that the two helicities are the same and that $\mathcal{Q}_S^+$ has conformal weights $h=\bar{h}=\frac{3}{2}$, we can write the two point function above as 
\begin{equation}
\langle \mathcal{N}(\vec{z}_i)\mathcal{N}(\vec{z}_j)\rangle = \int [d^2 w_i]\int [d^2 w_j] |z_i-w_i|^2|z_j-w_j|^2\langle{\rm{in}}|\mathcal{Q}_{S_1}^+(\vec{w}_1)\mathcal{Q}_{S_2}^+(\vec{w}_j)|{\rm{in}}\rangle
\end{equation}
where the integrals here are just the respective shadow transforms and $[dw^2]$ denotes the measure on the round sphere. These shadow transforms are particularly simple for the scalar modes, while the radiative modes can be reached by derivatives as in figure~\ref{fig:CN}. A similar expression exists for the one point function of a memory operator as it will just be the shadow of the soft one point function. Using~\eqref{eq:QS+} and~\eqref{eq:2ptconn} we can then start from our soft/hard charge relations and lift these to statements about the memory one- and two-point functions.

This procedure lets us co-opt ANEC correlators to get to the memory correlators, despite the fact that the phase space integrals for the `in-in' waveforms are naively different than the diagrams that would compute the ANEC correlators at the same order. Let us now turn to a discussion of some of the implications of our findings.

\section{Discussion}\label{sec:disc}

In our computations above, we have seen that the supertranslation Ward identities persist to the `in-in' formalism. Moreover, the `in-in' formalism constrains the soft charge correlators to be related to ANEC correlators. The celestial diamond framework then lets us lift these to memory correlators. We observed that this route is simpler than starting from the waveform computation and taking a soft limit, and that we can set up the computation in a manner that makes the the Ward identity manifest. This was seen in section~\ref{sec:softcharge} via matching the soft and hard contribution as well as by explicit computation. Namely we could use how the soft factors localize to contact terms to put the $\mathcal{Q}_S$ correlators on a similar footing in section~\ref{sec:softcharge} and then use shadows in section~\ref{sec:memmem} to return to the radiative modes. We expect this to also explain the manner in which analytic form of the waveforms differs from the ANEC correlators.

These discussions also raise the question of how well the celestial models for the soft sector~\cite{Himwich:2020rro,Kapec:2021eug,He:2024vlp} can capture statements about ANEC correlators. One distinction is that the form of the correlator depends on the in state we've prepared~\cite{Hofman:2008ar}. At the same time we see that the soft charge correlators should be $O(G^2)$ suppressed as compared to the ANEC correlators (which in a purely gravitational theory are also suppressed by powers of $G$~\cite{Herrmann:2024yai}), suggesting a tension with the scalings in~\cite{He:2024vlp} that requires a different explanation. We note that celestial correlators in in different backgrounds were considered in~\cite{Gonzo:2022tjm}. In the gauge theory case, some mileage has been made using these symmetries to constrain the all-orders structure of collinear limits of gluons in~\cite{Magnea:2021fvy}. It would be important to resolve if non-trivial higher point ANEC correlators point to more complicated infrared dynamics than the celestial models or can be simplified using insights from the asymptotic symmetries.

Natural next steps include generalizing to other gauge theories where the same infrared triangle structure is known to govern the relevant structure, as well as to the subleading soft graviton where we anticipate learning more about generalizations of the ANEC. We expect that the parallel discussion in gauge theories will follow directly from the results of this paper, in fact some of it has been done in~\cite{Gonzalez:2025ene} but that the next order in gravity will be more complex. It will also be extremely interesting to directly calculate higher point correlators of memory operators to directly verify the predictions derived in this paper.

\section*{Acknowledgments}

It is a pleasure to thank Matheus Balisa, Freddy Cachazo, Clifford Cheung, Hernan Gonzalez, Temple He, Prahar Mitra, Brett Oertel, Julio Parra-Martinez, Andrzej Pokraka, Jakob Salzer, and Sasha Zhiboedov for useful discussions. IM is supported by the DOE Early Career Award DE-SC0025581, and the Sloan Foundation. SN and SP are supported by the Celestial Holography Initiative at the Perimeter Institute for Theoretical Physics and the Simons Collaboration on Celestial Holography.  Research at the Perimeter Institute is supported by the Government of Canada through the Department of Innovation, Science and Industry Canada and by the Province of Ontario through the Ministry of Colleges and Universities.

\bibliographystyle{JHEP}
\bibliography{references}

\end{document}